\documentclass[journal]{IEEEtran}

\pdfoutput=1
\ifCLASSINFOpdf
\else
\fi

\usepackage[cmex10]{amsmath}
\usepackage{amssymb,amsthm,amsfonts}
\usepackage{stackrel,graphicx,subfig}
\usepackage{color}
\usepackage{epsfig,amssymb,mathrsfs}
\usepackage{soul,dsfont,enumerate}
\usepackage[export]{adjustbox}
\usepackage{booktabs}
\usepackage{multirow}

\newcommand{\shape}{\mathcal S}
\newcommand{\domain}{\Omega}
\newcommand{\digital}{\mathbf D}
\newcommand{\ud}{\, \mathrm{d}}
\newcommand{\consistency}{\mathcal C_\Omega(\digital;~f_1,...,f_{m^2})}

\newcommand{\R}{ \mathbb{R}}

\newcommand{\Lam}{\boldsymbol\lambda}

\theoremstyle{plain}\newtheorem{theo}{Theorem}
\theoremstyle{plain}\newtheorem{defi}{Definition}
\theoremstyle{plain}\newtheorem{lem}{Lemma}
\theoremstyle{plain}

\begin{document}

\title{Shapes From Pixels}

\author{Mitra~Fatemi,
		Arash~Amini,
        Loic~Baboulaz,
        and~Martin~Vetterli
\thanks{M. Fatemi, L. Baboulaz and M. Vetterli are with the Department
of Computer and Communication Sciences, \'Ecole Polytechnique F\'ed\'eral de Lausanne (e-mail: \{mitra.fatemi, loic.baboulaz, martin.vetterli\}@epfl.ch).}
\thanks{A. Amini is with the Department of Electrical Engineering, Sharif University of Technology (e-mail: aamini@sharif.edu).}}

\maketitle

\begin{abstract}
Continuous-domain visual signals are usually captured as discrete (digital) images. This operation is not invertible in general, in the sense that the continuous-domain signal cannot be exactly reconstructed based on the discrete image, unless it satisfies certain constraints (\emph{e.g.}, bandlimitedness). In this paper, we study the problem of recovering shape images with smooth boundaries from a set of samples. Thus, the reconstructed image is constrained to regenerate the same samples (consistency), as well as forming a shape (bilevel) image. We initially formulate the reconstruction technique by minimizing the shape perimeter over the set of consistent binary shapes. Next, we relax the non-convex shape constraint to transform the problem into minimizing the total variation over consistent non-negative-valued images. We also introduce a requirement (called reducibility) that guarantees equivalence between the two problems. We illustrate that the reducibility property effectively sets a requirement on the minimum sampling density. One can draw analogy between the reducibility property and the so-called restricted isometry property (RIP) in compressed sensing which establishes the equivalence of the $\ell_0$ minimization with the relaxed $\ell_1$ minimization. We also evaluate the performance of the relaxed alternative in various numerical experiments.

\end{abstract}


%
\IEEEpeerreviewmaketitle

\section{Introduction}\label{sec:intro}
\begin{figure}
\centering
\includegraphics[width=0.8\linewidth]{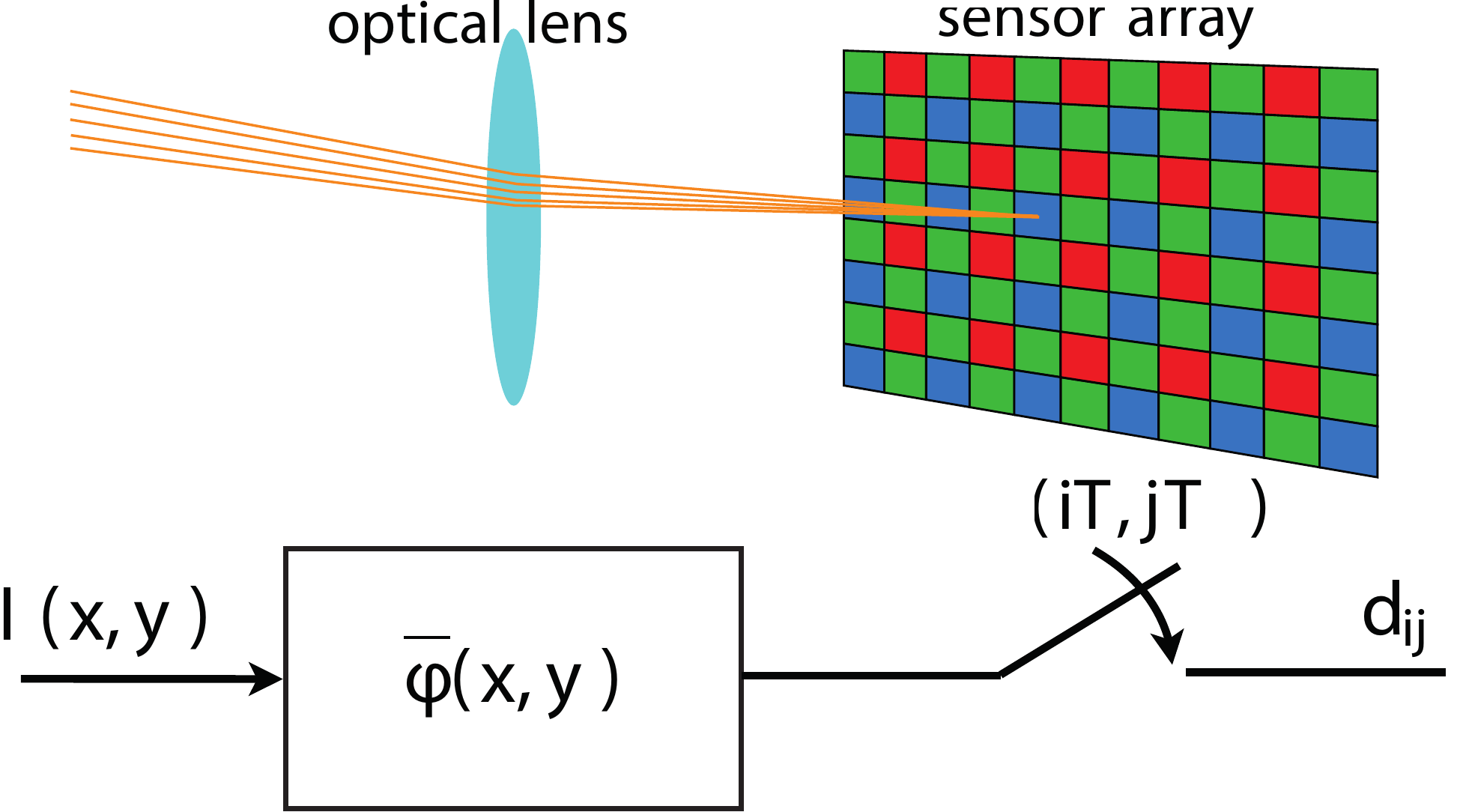}
\caption{Sampling system implemented by a digital camera; the effect of the optical lens and the sensor array on 2D visual signals can be modeled by filtering followed by sampling in space.}\label{fig:sampling_system}
\end{figure}

Sampling is at the heart of all digital signal acquisition devices. To store and process the data, we need to convert continuous-domain signals into a sequence of numbers. Since the continuous-domain signal will no longer be available after this stage, we expect the sequence to provide an exact or at least a fair representation of this signal. The Shannon sampling theory and its variations consider sampling strategies for the class of signals living in a shift-invariant space. Interested readers are referred to \cite{Jerri77,Unser2000,Vetterli-Book2014} for a comprehensive study of the topic. Here, we focus on the imaging application carried out by a digital camera, in which the optical lens and the sensor array are responsible for the sampling procedure (Figure \ref{fig:sampling_system}). The physics of the device imply that the measured sensor values---pixel values of the digital image---are samples of a filtered visual signal. The impulse response of the involved filter is called the point spread function (PSF) and is determined by the optical system. On the other hand, the sensor array controls the number of samples and the sampling resolution.

%

\begin{figure}[t]
\centering
\includegraphics[width=0.3\linewidth]{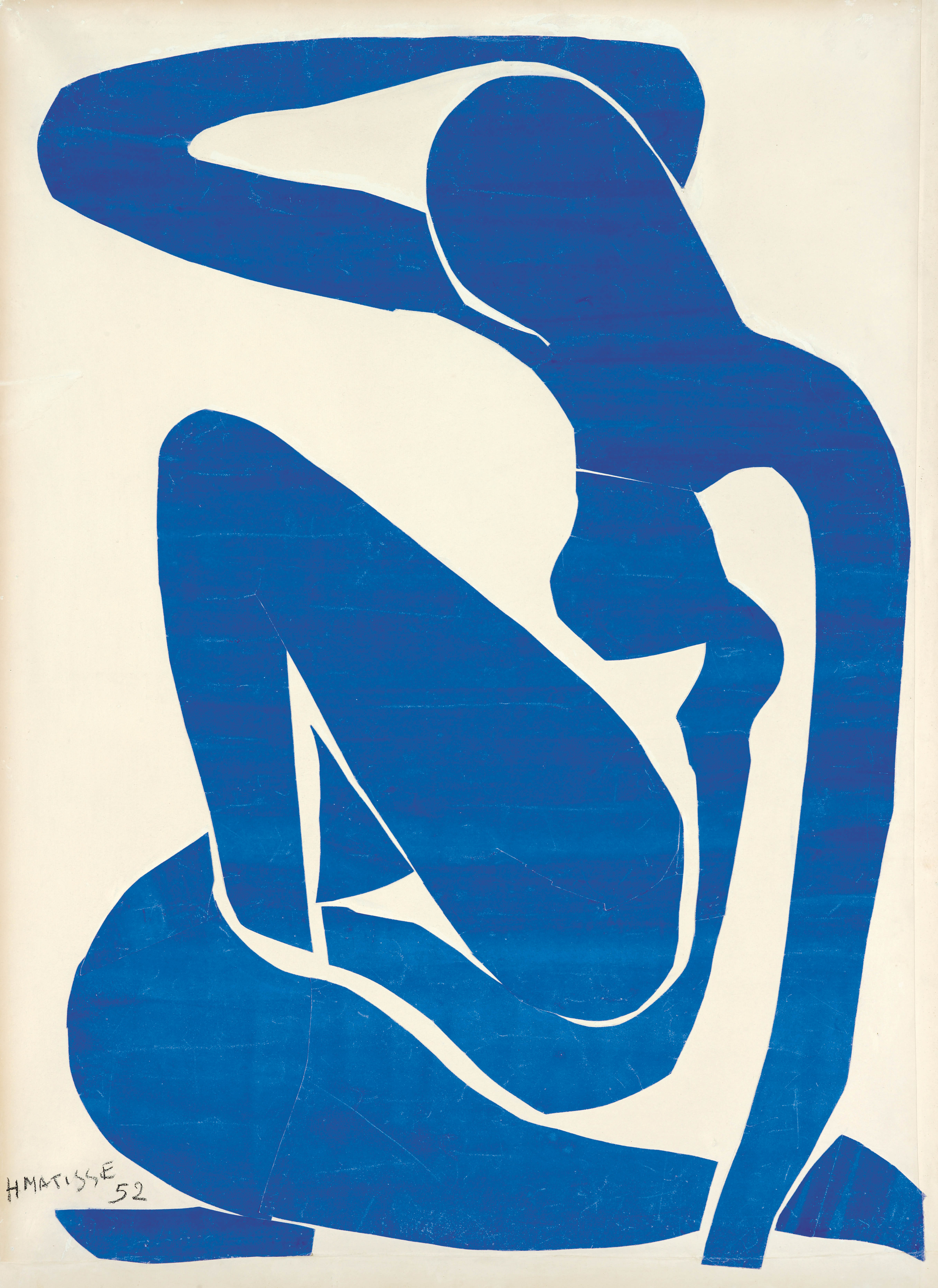}
\caption{Cutout by Henri Matisse (1952) that can be classified as a shape image.}
\label{fig:Matisse}
\end{figure}

In this paper, we study the problem of recovering continuous-domain visual signals from their samples. Within the broad range of visual signals, we shall focus only on the class of \emph{shape} signals over a fixed background. A shape is mathematically described as the characteristic function of a union of a few connected subsets. We call these signals \emph{shape images}. We allow for shapes with arbitrary geometries as long as they have smooth boundaries.
Examples of shape images can be found among artworks such as woodcut prints, cutouts and lithographs (Figure \ref{fig:Matisse}).

The shape images have only two different intensity values (0 and 1). However, the filtering effect caused by the PSF smooths out  sharp intensity transitions and results in measurements with varied intensity levels. In short, the sampling process projects  binary shape images onto gray-scale discrete images (Figure \ref{fig:jar}). 
In this paper, we are interested in a strategy to recover the original binary image or a binary approximation thereof from measurements. In any case, the recovered binary image should be able to regenerate the same measurements. This requirement assures us that we cannot discriminate between the original and the recovered images at least at the output of the sampling block. A reconstruction of the original image that satisfies this condition is called \emph{measurement-consistent} or \emph{consistent} for short \cite{Thao94, Unser94, Unser98}. The problem of consistent shape image reconstruction appears in applications where the aim is to exactly locate or describe the objects in a scene; astronomical imaging, quality monitoring in manufacturing, biomedical imaging and high-quality artwork rendering are a few examples. 

\subsection{Related Works}
Due to the diverse shape geometries and sharp intensity transitions on the boundaries, this class of visual signals, like many other real-world signals, are neither bandlimitted nor belong to a shift invariant subspace. Hence, the classical sampling results do not apply here. A similar scenario happens for the class of 1D signals studied in \cite{Vetterli2002, Blu2008},  known as signals with finite rate of innovation (FRI). It is shown that the discrete samples can lead to perfect signal recovery, although the signals are not necessarily  bandlimitted. A generalization to 2D FRI signals is presented in \cite{Maravic2004,Shukla2007,Chen2012}, with the goal of recovering convex polygonal shapes from the gray-scale pixels. A different approach is devised in \cite{Pan2014} by considering the boundary curves in a shape image as the zero-level-sets of specific 2D FRI signals. Due to the FRI requirements, exact recovery relies on the PSF satisfying the so-called Strang-Fix condition. Furthermore, the FRI model admits limited shape geometries.

The shape image recovery can also be viewed as fitting boundary curves to the interpolated gray-scale image (high-resolution version of the measurements). Such methods are widely known as segmentation techniques that fit deformable curves to gray-scale images, and include active contour algorithms also known as snakes. Based on the curve models, they are classified as point snakes \cite{Kass87}, geodesic snakes \cite{Malladi95,Caselles97} and parametric snakes \cite{Figueiredo2000,Brigger2000}. In all cases, the segmentation algorithm is formulated by minimizing a snake energy functional that depends on the gray-scale image and the model of the boundary curves.  However, it does not take the PSF into account \cite{Gonzalo2015}. As a consequence, the resulting binary image is likely to fail the consistency requirements.

%

\subsection{Contributions}

In this work, we propose a method to recover a measurement-consistent shape image that has continuously twice differentiable ($C^2$) boundary curves. Our approach is direct in the sense that it avoids intermediate curve fitting steps, and finds the shape with minimum perimeter. We formulate the method as an optimization problem constrained by the measurements (i.e. pixels), where the functional is the continuous-domain total variation (TV). Ideally, we should restrict the search domain to binary images. This leads to a non-convex problem which is computationally intractable. Hence, we consider the convex relaxation in which the search is over the set of all non-negative-valued images. Under a minimum resolution requirements (see Definition \ref{defi:unifiable} for an explicit explanation), we prove that all the solutions to the non-convex problem are  minimizers of the convex relaxation (see Theorem \ref{theo:ConsistentShape}). 


The number of constraints in the convex problem equals the number of pixels. However, we demonstrate that when the resolution requirement is satisfied, the multiple constraints can be replaced with a single one formed by a wisely chosen linear combination of them. This reduces the problem to an equivalent TV minimization problem with a single constraint, which is known in the literature as the \emph{Generalized Cheeger problem} \cite{Ionescu2005}. A generalized Cheeger set is a shape with minimum perimeter and a fixed weighted integral. This equivalence allows us to apply the existing results that show the Cheeger solutions are among the minimizers of the relaxed problem \cite{Ionescu2005,Carlier2007}.


The advantage of our method compared to the FRI works in \cite{Maravic2004}-\cite{Pan2014} is that, we do not constrain the boundary curves by any specific model. Instead, we let the sampling kernel and the measurement values decide for them. As a result, there is less restriction on the achievable shape geometries. Besides, the choice of the PSF is arbitrary, and does not need to satisfy the Strang-Fix condition.

\begin{figure}[t]
\centering
\subfloat[]{\includegraphics[width=0.4\linewidth]{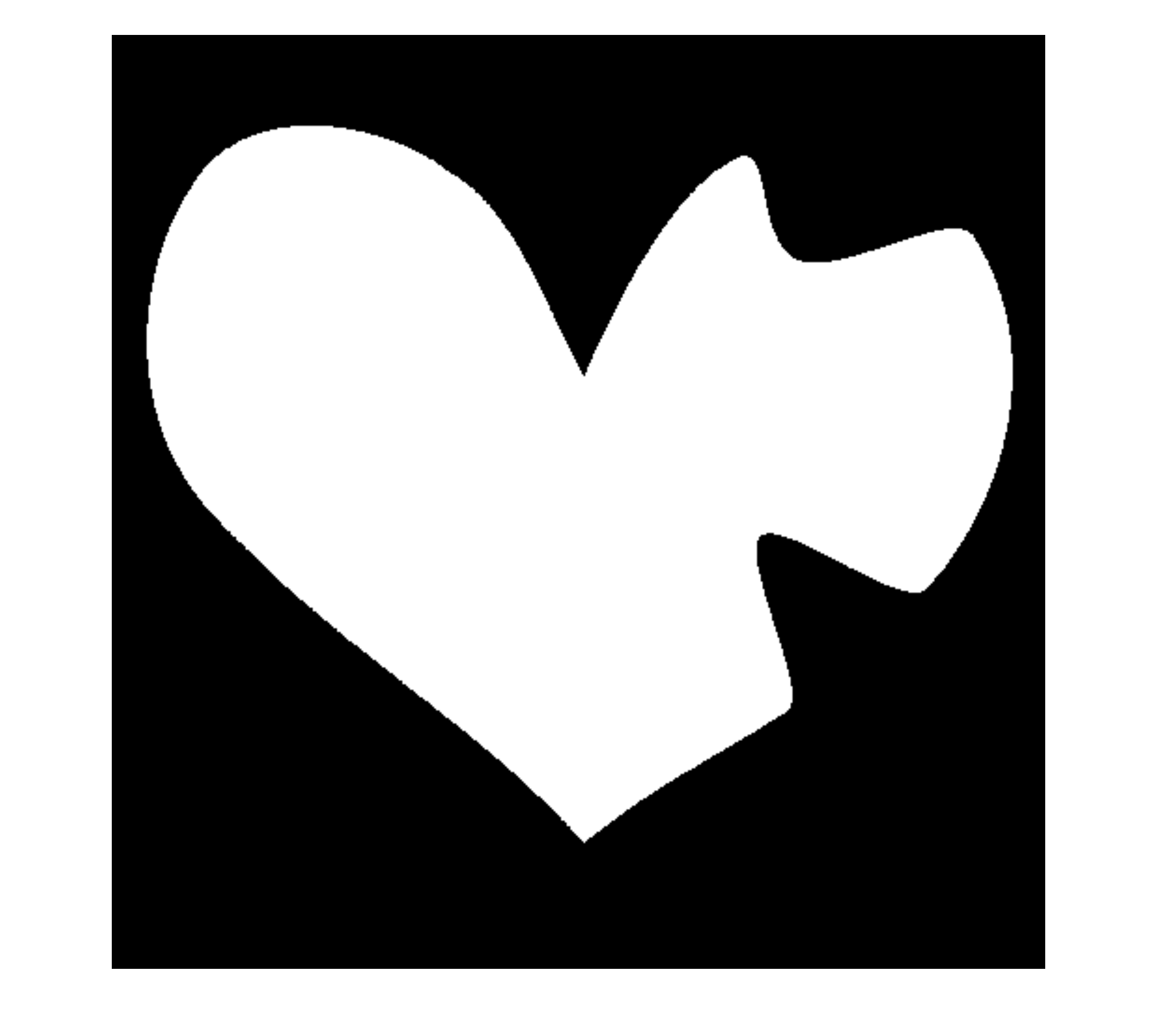}%
\label{fig:cont_jar}}
\subfloat[]{\includegraphics[width=0.4\linewidth]{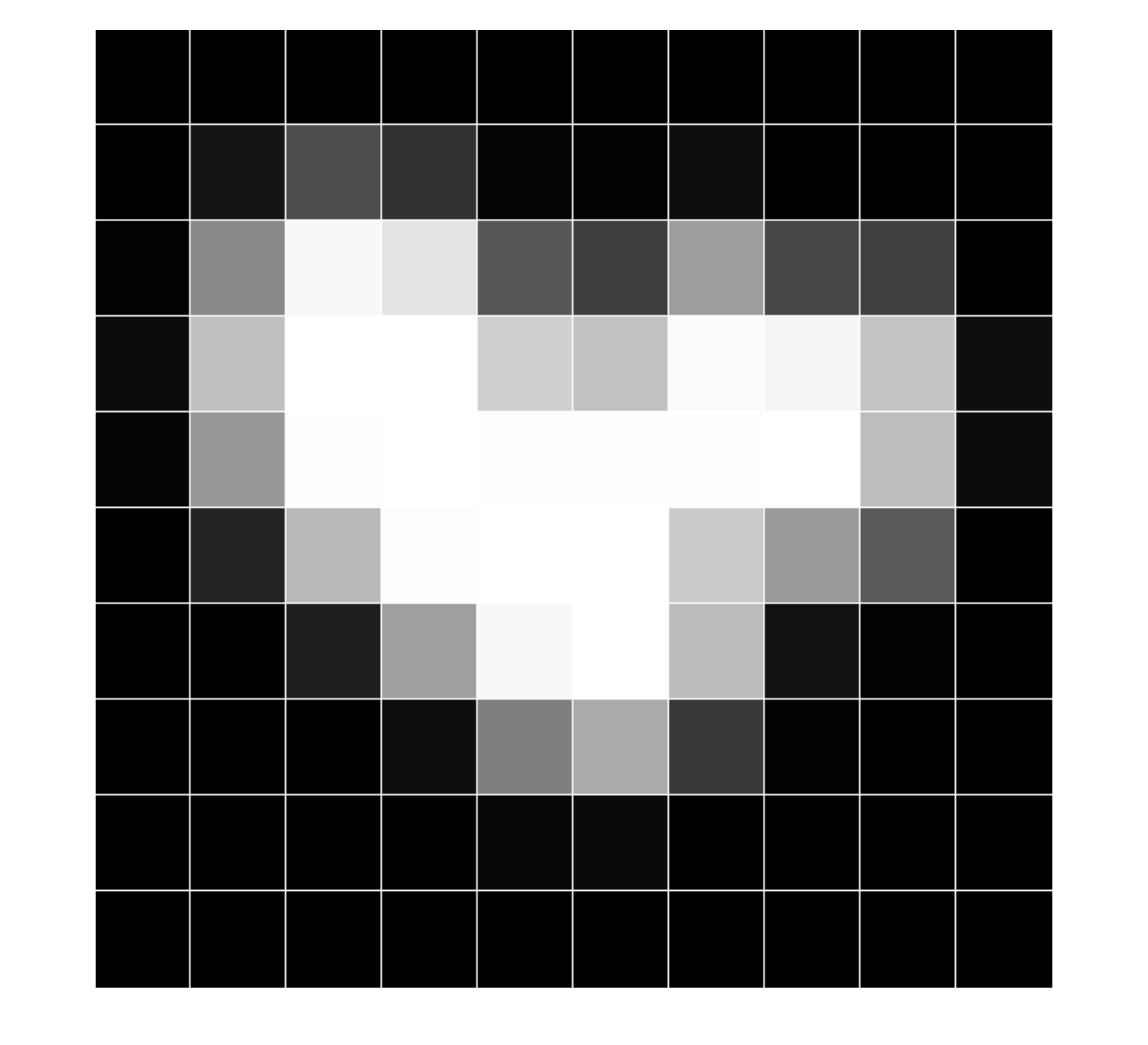}%
\label{fig:dis_jar}}
\hfill
\caption{Example of a bilevel image and its acquisition: (a) a shape image, and (b) pictorial representation of $10\times10$ measurements, generated with a bilinear B-spline sampling kernel.}
\label{fig:jar}
\end{figure}

\subsection{Organization of the paper}
We explicitly define the problem and the used notations in Section \ref{sec:problem}. We continue by reviewing the concept of Cheeger sets and the existing results in Section \ref{sec:Cheeger}. 
In Section \ref{sec:algorithm}, we present the theoretical results. We employ the primal-dual algorithm of \cite{Strekalovskiy2014} for the numerical approximation of the solutions to the convex minimization problem in Section \ref{sec:implementation}. This algorithm enables us to study the performance of the proposed shape recovery method in Section \ref{sec:algorithm} through numerical experiments. Finally, we conclude the paper in Section \ref{sec:conclusion}.

\section{Problem Definition}\label{sec:problem}
\begin{figure*}[!t]
\centering
\subfloat[]{\includegraphics[width=0.16\linewidth,valign = c]{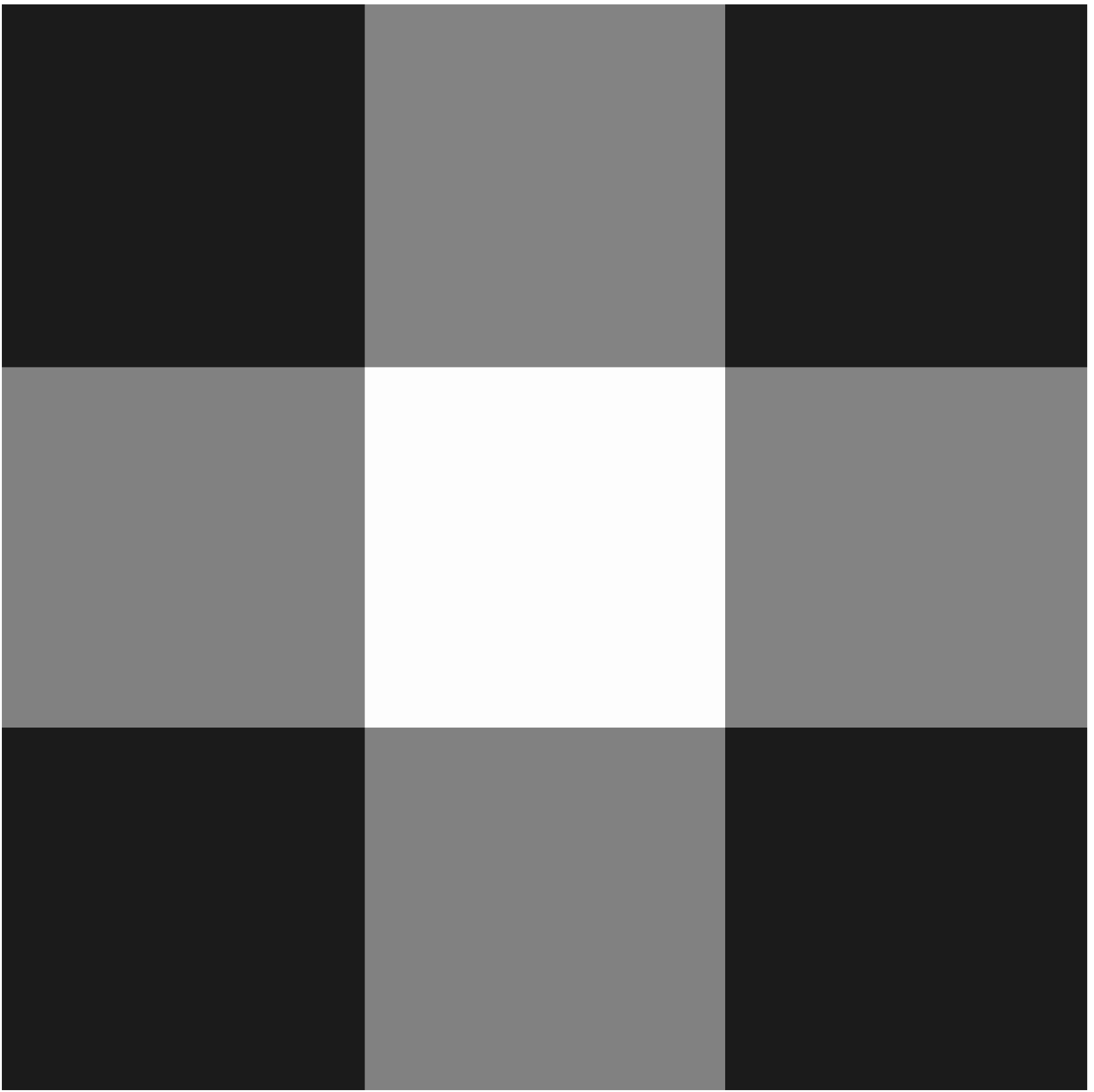}}
\hspace{20pt}
\subfloat[]{\includegraphics[width=0.15\linewidth,valign = c]{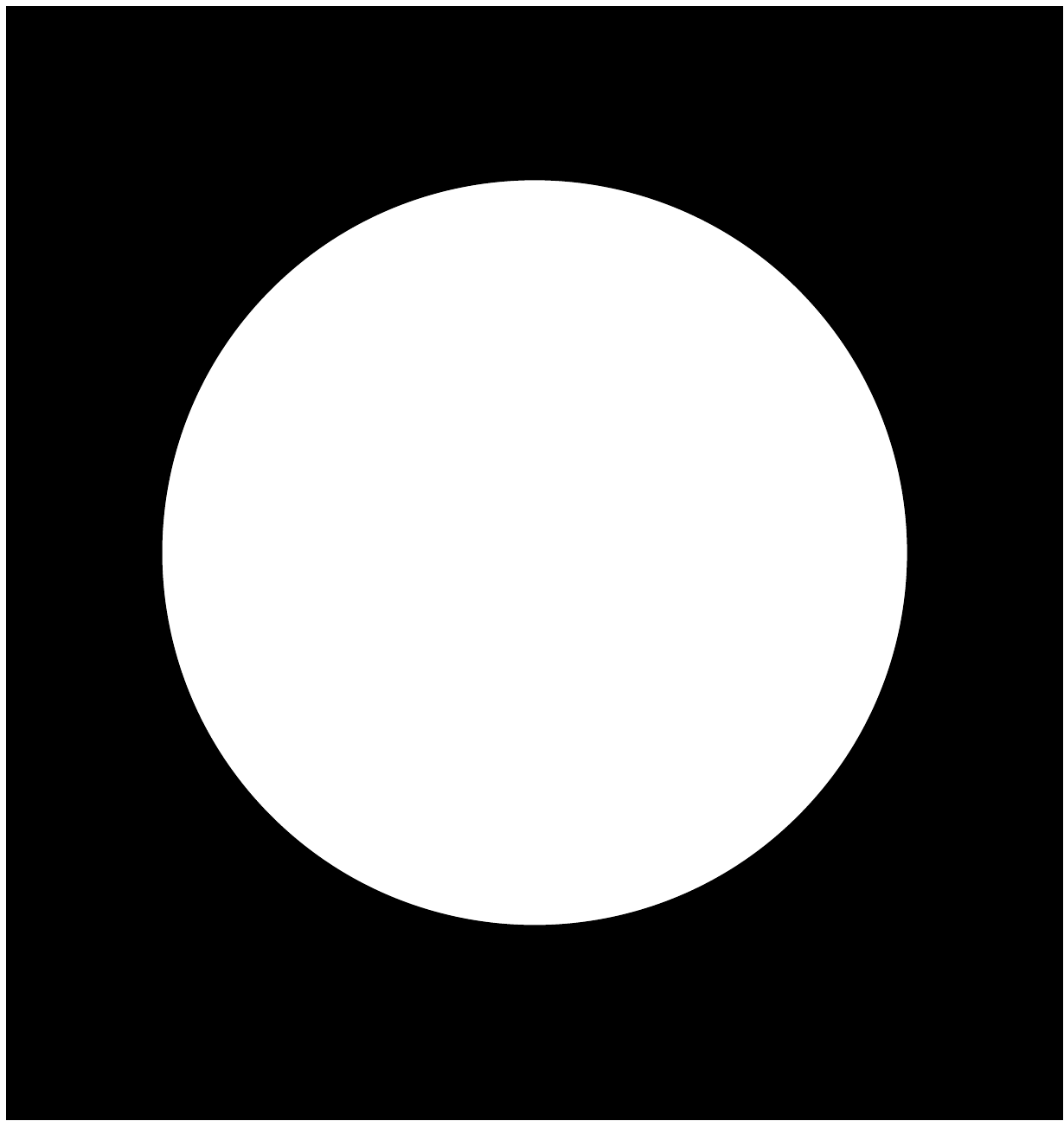}}
\hspace{20pt}
\subfloat[]{\includegraphics[width=0.15\linewidth,valign = c]{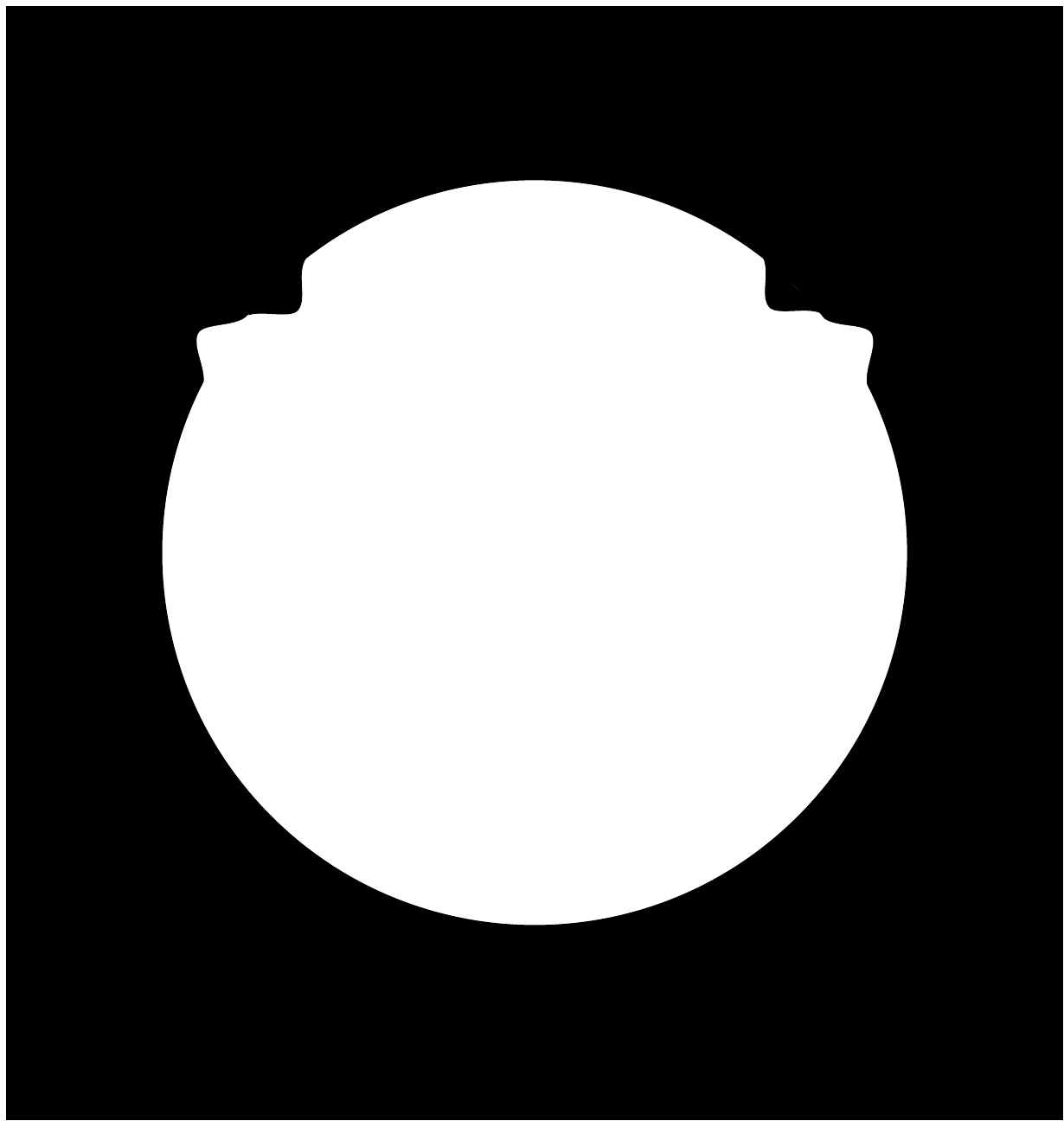}}
\hspace{20pt}
\subfloat[]{\includegraphics[width=0.15\linewidth,valign = c]{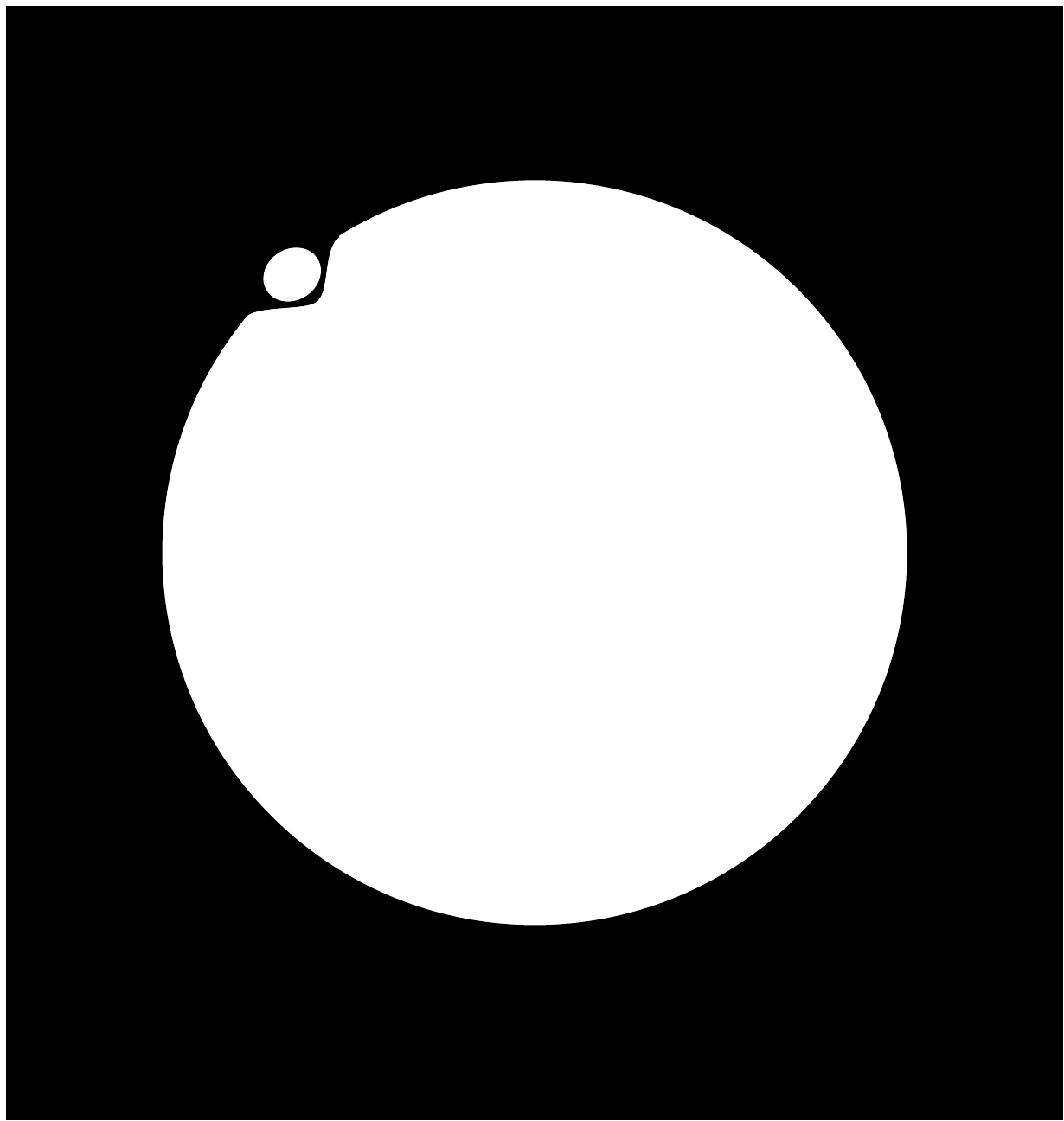}}
\caption{Perimeter minimization prevents unnecessary details and extra connected components to appear in the shape. In this figure, all shapes are consistent with the measurements in (a) but the shapes in (c) and (d) have higher perimeters due to extra details on the boundary and additional component, respectively.}
\label{fig:least_perimeter}
\end{figure*}

We denote by $I(x,y)$ the continuous-domain image with domain $\domain=[0,1]^2$. Also, we represent the discrete (or measurement) image  by $\digital$. When $\digital$ is the output of an $m\times m$-pixel digital camera with PSF $\overline\phi(x,y)=\phi(-x,-y)$, we can relate the $m^2$ pixels $d_{ij},~1\leq i,j\leq m$ of $\digital$ to the image $I(x,y)$ as
\begin{align}\label{eq:pixels}
d_{ij}&=\frac{1}{T^2}~\overline\phi(\frac{x}{T},\frac{y}{T})~*~I(x,y)~\arrowvert_{(x,y)=(jT,iT)}\nonumber\\
&=\iint_\Omega\frac{1}{T^2}\phi(\frac{x}{T}-j,\frac{y}{T}-i)~I(x,y)\ud x\ud y,
\end{align}
where $T$ is the sampling period (Figure \ref{fig:sampling_system}).
%

In the consistent image recovery problem, we wish to find an approximation $\tilde I$ of the original image that regenerates the same measurement pixels. This means that the reconstruction error $I-\tilde I$ between the original image and its approximation is in the null space of the imaging process. Equivalently, the two images are perceived as identical by the imaging device. Let $k=(j-1)m+i,~1\leq k\leq m^2$ represent the equivalent index of $d_{ij}$ in the vertical raster scan of $\digital$. Also, let $$f_k(x,y) = \frac{1}{T^2}\phi\Big(\frac{x}{T}-\lceil k/m\rceil~,~\frac{y}{T}-((k\mod m) +1)\Big)$$ indicate the sampling kernel in \eqref{eq:pixels} associated with $d_k$. We denote by $\consistency$ the set of all non-negative-valued images over the domain $\Omega$ that are consistent with $\digital=[d_k]_{1\leq k\leq m^2}$,
\begin{align}\label{eq:consistency}
&\consistency\\
&=\Big\{I\in BV(\Omega), I\geq0~;~\iint_\Omega If_k\ud x\ud y = d_k,~ {\scriptstyle 1\leq k\leq m^2}\Big\}.\nonumber
\end{align}
Here, $BV(\Omega)$ stands for the set of functions over $\Omega$ with bounded variation; {i.e.}, all elements of $BV(\Omega)$ have well-defined and finite total variation values.



Consistent image recovery is equivalent to finding an element of $\consistency$. In the consistent shape recovery problem, we limit the permissible solutions to the shape characteristic functions.
Let $\shape$ be a subset of $\domain$. We denote by $\chi_\shape(x,y)$ the characteristic function of $\shape$ over $\domain$
\begin{eqnarray*}\label{eq:indicator}
\chi_\shape(x,y)=\begin{cases}
1,& \text{if }(x,y)\in\shape\\
0,&\text{if }(x,y)\in\Omega\setminus\shape.
\end{cases}
\end{eqnarray*}
We call $\shape$ a shape if it is the union of a finite number of connected subsets of $\Omega$. In this case, we call $\chi_\shape$  a shape image.

We assume that the sampling kernel in \eqref{eq:pixels} is always positive and has unit $\ell_1$ norm
$$\iint\phi(x,y)\ud x\ud y=1.$$
Consequently, the pixels of the discrete image associated with a shape characteristic function shall take continuous values in the range $[0,1]$. An example of a shape characteristic function and its associated $10\times 10$ discrete image is shown in Figure \ref{fig:jar}. We recall that Figure \ref{fig:dis_jar} provides a pictorial representation of the 100 pixels in $\digital$ and it should not be mistaken with a piecewise constant approximation of the original image.

The consistent shape reconstruction problem is equivalent to finding a shape image $I=\chi_{\shape}(x,y)\in\consistency$ for the set of $m^2$ pixels $0\leq d_k\leq 1$ in $\digital$. Among all possible candidates, we are interested in shape images with minimum perimeter. This way we reject shapes with extra connected components and excessive boundary details (see Figure \ref{fig:least_perimeter}). 

Minimum-perimeter consistent shapes are the global minimizers of the following problem
\begin{align}\label{eq:P_0}
\tag{$P_0$}
&\inf_{\substack{\shape\subset\Omega,\,\, \chi_\shape\in{BV}}}~\text{Per}(\shape),&\\
&\textstyle{\text{s.t.}~I=\chi_\shape\in\consistency,}\nonumber
\end{align}
where $\text{Per}(\shape)$ is the perimeter of $\shape$. Problem \eqref{eq:P_0} is a variational non-convex problem and it is prone to having many local minima. This makes it very likely that common gradient descent methods get trapped in local minima. While in problems of this sort, global minimizers are usually all reasonable solutions, the local minima can be blatantly false. In the next sections, we show that if the discrete image $\digital$ satisfies a resolution requirement defined in Definition \ref{defi:unifiable}, the minimum-perimeter consistent shapes are the minimizers of a convex relaxation of \eqref{eq:P_0}. Furthermore, we conjecture that under this condition, there is a unique minimum-perimeter consistent shape which is also the unique solution of the convex problem. In the experimental section, we present an algorithm for the recovery of this solution.

%

\section{Cheeger Sets}\label{sec:Cheeger}
An image is called consistent with the measurements if it complies with all the constraints in \eqref{eq:P_0}. Essentially, each pixel of the discrete image enters \eqref{eq:P_0} as a constraint, resulting in an optimization with many constraints. In addition, we are also restricting the search domain to bilevel images, which further complicates the minimization task. The simplest scenario of having only one single pixel (measurement) is a well-studied topic known as the Cheeger problem. There is already a rich literature regarding the existence, uniqueness properties, regularity (smoothness) of the boundary and numerical evaluation of such sets for almost arbitrary kernels $f$. In this section, we  present a brief review of the Cheeger problem and related results upon which we build our general multi-constraint minimization problem. The details for the latter will be discussed in the next section.  

The Cheeger problem can be directly extended to higher dimensions; however, for the purpose of image recovery, we focus on 2D signals in this paper. Let $\Omega$ be a subset of $\R^2$. 
The Cheeger sets of $\Omega$ are defined as those $\shape\subset\Omega$ that minimize the ratio of the perimeter over the area,
\begin{align}\label{eq:Cheeger}
\tfrac{\text{Per}(\shape)}{\iint_\shape\ud x\ud y}.
\end{align}

It is common to represent $\text{Per}(\shape)$ in terms of the total variation of the shape image $\chi_\shape$. For this purpose we invoke the \textit{coarea} formula that for a positive function $u(x,y):\Omega\rightarrow\R^{\geq0}$ implies that
\begin{align*}
TV(u)=\iint_\Omega|\nabla u|\ud x\ud y=\int_{0}^\infty \text{Per}\big(E(u;\mu)\big)\ud\mu,
\end{align*}
where 
$$E(u;\mu) =\{(x,y)\in\Omega~\Big|~u(x,y)\geq\mu\}$$ are the level-sets of $u(x,y)$. This immediately indicates that $\text{Per}(\shape)=TV(\chi_\shape)$.

\begin{figure}
\centering
\subfloat[]{\includegraphics[width=0.45\linewidth,valign=m]{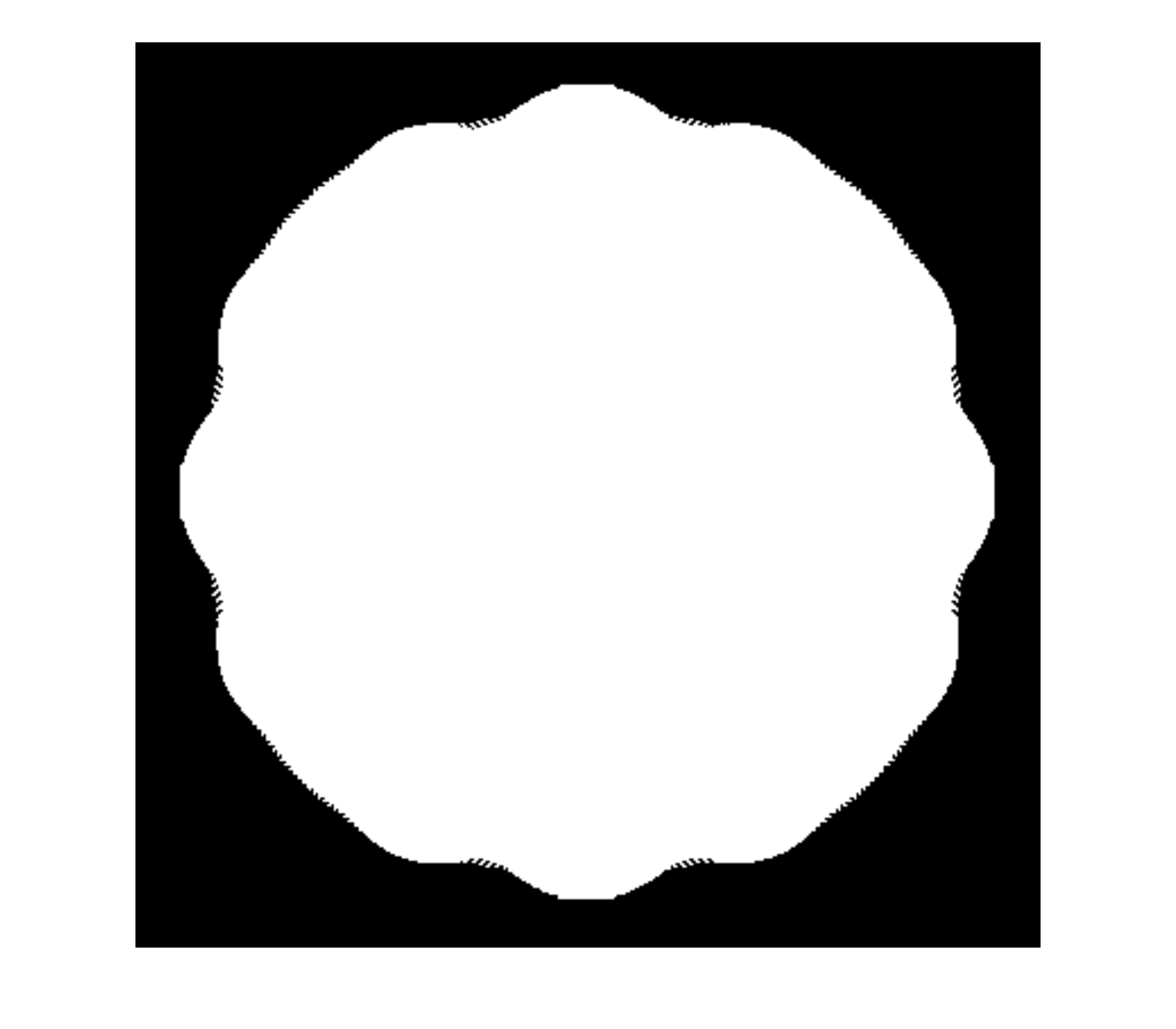}%
\label{fig:CheegerSet}}
\subfloat[]{\includegraphics[width=0.45\linewidth,valign=m]{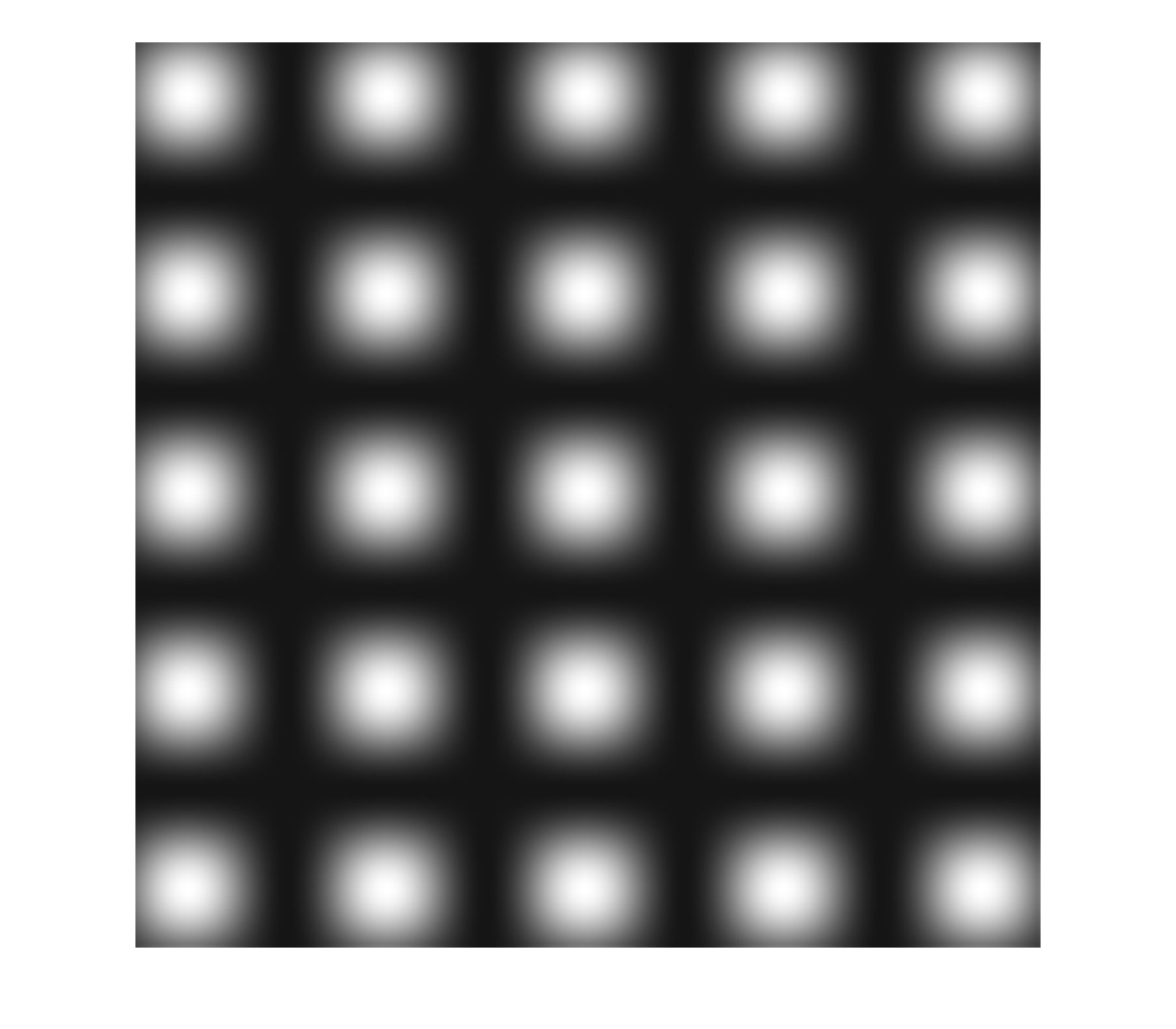}%
\label{fig:Cheeger_kernel}}
\hfill
\caption{Example of (a) a generalized Cheeger set, for the weight kernels $g\equiv1$ and $f$ as displayed in (b).}
\label{fig:Cheeger}
\end{figure}

We can expand the definition of a Cheeger set by introducing two weight kernels in the nominator and denominator of \eqref{eq:Cheeger} \cite{Ionescu2005}. 
Indeed, a generalized Cheeger set is a shape minimizer of 
%
%
\begin{equation}\label{eq:GenCheeger}
\inf_{\substack{\shape\subset\Omega,\,\, \chi_\shape\in{BV}}}\frac{\iint_\Omega g|\nabla \chi_\shape|\ud x\ud y}{\iint_\Omega f\chi_\shape\ud x\ud y}.
\end{equation}
Note that for $f=g\equiv 1$ we obtain the standard Cheeger sets. For a simple domain such as a square, the Cheeger set is unique and has a certain shape but depending on the choice of the weight kernels, generalized Cheeger sets can have very diverse shapes. Figure \ref{fig:Cheeger} displays an example of a generalized Cheeger set for $g\equiv 1$ and $f$ as in Figure \ref{fig:Cheeger_kernel}.

Except for special choices of $f$ and $g$, the minimizer of \eqref{eq:GenCheeger} shall not be unique.
Furthermore, the minimization in \eqref{eq:GenCheeger} is over a non-convex set, which is computationally intractable in general. However, an interesting result by Strang in \cite{Strang83} (see also \cite{Carlier2007}) shows that all global minimizers of \eqref{eq:GenCheeger} (all Cheeger sets) are the level-sets of the solution(s) to 
\begin{equation*}
\inf_{I:\Omega\rightarrow\R^{\geq0}}\frac{\iint_\Omega g|\nabla I|\ud x\ud y}{\iint_\Omega fI\ud x\ud y}.
\end{equation*}
Note that the search domain in the latter problem consists of all non-negative-valued images (not necessarily bilevel), which is a convex set. The following statement of this result by \cite{Carlier2007} is more aligned with our approach in the next section.
\begin{theo}[\cite{Carlier2007}]\label{theo:Cheeger}
Let $I$ be a minimizer of 
\begin{equation}\label{eq:CheegerConv}
\inf_{I\in BV_\Omega(f)}\iint_\Omega g|\nabla I|\ud x\ud y,
\end{equation} 
where
\begin{align*}
BV_\Omega(f)=\Big\{I\in BV(\Omega),~I\geq0~;~\iint_\Omega fI\ud x\ud y = 1\Big\}.
\end{align*}
Then, for every $\mu\geq 0$ such that the level-set $E(I;\mu)$ is nonempty, 
\begin{align}\label{eq:bin_sol}
\frac{1}{\iint_{E(I;\mu)}f\ud x\ud y}\chi_{E(I;\mu)}
\end{align} 
is also a minimizer of \eqref{eq:CheegerConv}. 
\end{theo}

In a nutshell, Theorem \ref{theo:Cheeger} states that the minimizer set of \eqref{eq:CheegerConv} is closed under level-set evaluation; i.e., normalized (scaled) non-empty level-sets of a minimizer also belong to the set of minimizers. This helps in finding a bilevel solution to \eqref{eq:GenCheeger}, as finding any minimizer of the convex problem \eqref{eq:CheegerConv} necessarily leads to (at least) a bilevel image.

Another result proved in \cite{Carlier2007} indicates that the Cheeger sets are closed under set union. This immediately establishes the existence of a unique maximal Cheeger set that contains all the other ones \cite{Buttazzo2007}. Thus, we can remove the inherent ambiguity caused by non-uniqueness of the solutions to \eqref{eq:CheegerConv} by searching for the maximal set. 
However, finding the maximal set is not generally easy by considering the minimization of \eqref{eq:CheegerConv}. A regularization technique is proposed in \cite{Buttazzo2007} that applies asymptotically vanishing penalty terms to the cost function \eqref{eq:CheegerConv} and achieves the maximal set at the limit of the minimizers. Based on this idea, a numerical method is introduced in \cite{Carlier2009} that approximates the maximal Cheeger set on a finite grid. The method is robust to discretization as the approximations converge point-wise to the continuous-domain Cheeger set when the grid resolution increases.

As a final note, we discuss the influence of the weight kernels $f$ and $g$. In fact, Cheeger sets consist of smooth $C^2$ boundaries, irrespective of the choice of $f$ and $g$ \cite{Ionescu2005}. Nevertheless, it is known that the curvature of the boundaries is tightly controlled by these weight kernels. Formally, at each boundary point we have that \cite{Ionescu2005}
\begin{equation}\label{eq:curvature_bound}
|\kappa|\leq\frac{\mathcal{J}(\shape)\sup f+\sup\|\nabla g\|}{\inf g}
\end{equation}
where $\kappa$ stands for the curvature and $\mathcal J(\shape)$ is the cost value of the Cheeger set determined by the ratio in \eqref{eq:GenCheeger}. As we set $g\equiv 1$ in the rest of the paper, the effective bound on the curvature simplifies to $|\kappa|\leq\mathcal J(\shape)\sup f$. We will just briefly comment on employing a non-constant weight kernel $g$ in Section \ref{sec:conclusion}.

\section{Consistent Shape Recovery}\label{sec:algorithm}
Let us consider the problem \eqref{eq:P_0} for the case where the measurement image $\digital=[d_k]_{1\leq k\leq m^2
}$ consists of more than one pixel. Similar to the single-measurement setting, non-convexity of the problem is a computational barrier. Therefore, we opt to use a convex relaxation in the form of 
%
\begin{align}\label{eq:P_1}
\tag{$P_1$}
&\inf_{I\in\consistency}~\iint_\Omega|\nabla I|\ud x\ud y.&
\end{align}
By extending the search domain from binary (bilevel) shapes to all non-negative-valued images, the problem becomes convex. 
Nevertheless, due to multiple measurement constraints, this scenario obviously deviates from the conventional Cheeger problem.
%

In Theorem \ref{theo:ConsistentShape} we show that under certain conditions, the minimization in \eqref{eq:P_1} constrained by multiple measurements can be replaced with a similar minimization subject to a single constraint; i.e., we prove that \eqref{eq:P_1} could potentially have an equivalent  Cheeger problem. In fact, we use a wisely chosen linear combination of the measurements as the single measurement. The interpretation of \eqref{eq:P_1} in form of a Cheeger problem automatically implies the existence of a bilevel minimizer (\emph{e.g.}, the maximal Cheeger set) for \eqref{eq:P_1}. Thus, all shape minimizers of \eqref{eq:P_0} are also minimizers of the relaxed problem \eqref{eq:P_1}. Further, it proves the existence and uniqueness of a maximal consistent shape. In Theorem \ref{theorem:level-set}, we provide a simple test to verify whether a minimizer of \eqref{eq:P_1} is the maximal shape. This helps us to validate a numerical solution obtained via minimizing \eqref{eq:P_1}---which might not have a unique minimizer---as a binary consistent shape.   

The  mathematical requirements for the equivalence of \eqref{eq:P_1} with a Cheeger problem (existence of a suitable linear combination of the measurements) is stated in Definition \ref{defi:unifiable}; essentially these requirements imply that the sampling density used for obtaining the measurement image needs to be fine enough.

\subsection{Theoretical Results}\label{subsec:results}
We start by defining the maximal consistent shape.
\begin{defi}\label{defi:maximal-shape}
A maximal consistent shape with minimum perimeter, or MCSMP in short, is a solution to \eqref{eq:P_0} whose support contains the support of all other minimizers of \eqref{eq:P_0}. 
\end{defi}
Note that a MCSMP does not always exist. In general, the support union of two minimizers of \eqref{eq:P_0} does not necessarily generate a minimizer by scaling. It is evident by this fact that the claimed equivalent Cheeger problem plays a crucial role in our results. In Definition \ref{defi:unifiable} below we will describe the sufficient conditions that enable us to associate \eqref{eq:P_0} or \eqref{eq:P_1} to a Cheeger problem. 

Before stating Definition \ref{defi:unifiable}, we introduce a few notations used in the rest of this section. As we need to linearly combine the measurement constraints, we represent the $n$-dimensional coefficient set for the convex combinations by $\Delta_n$: 
\begin{align*}
\Delta_n \triangleq \Big\{ (\lambda_1,\dots,\lambda_n)\in\R^n ~\Big|~ 0\leq \lambda_i,\, \sum_{i=1}^{n}\lambda_i=1\Big\}.
\end{align*}

For non-negative-valued images, a zero measurement can only happen when the image vanishes over the support of the corresponding sampling kernel. Thus, we can exclude the support region from our search domain. 
\begin{defi}\label{def:Reduced_Domain}
For the measurements $\digital=[d_k]_{1\leq k\leq m^2}$ corresponding to the pixels $0\leq d_k\leq 1$ and sampling kernels $f_1,\dots,f_{m^2}$, let $\rho$ denote the number of non-zero pixels and
\begin{align*}
A = \{i\,\big|\, d_i>0\} = \{a_1,\dots,a_{\rho}\}
\end{align*}
stand for the index set of active pixels. 
We define the \emph{reduced domain} $\Omega_{\rm r}$ by
\begin{align*}
\Omega_{\rm r}=\Omega_{\rm r}(\digital;f_1,\dots,f_{m^2}) = \Omega\setminus \, \underset{\scalebox{0.7}{$i\in\{1,\dots,m^2\}\setminus A$}}{\scalebox{1.4}{$\cup$}} \text{supp}(f_i).
\end{align*}
\end{defi}

\begin{defi}\label{def:Reduced_Kernel}
For the measurements $\digital=[d_k]_{1\leq k\leq m^2}$, sampling kernels $f_1,\dots,f_{m^2}$, and a vector $\boldsymbol{\lambda}\in\Delta_{\rho}$, we define the \emph{reduced kernel} $f^{\boldsymbol{\lambda}}:\Omega_{\rm r}\mapsto \R^{\geq0}$ by
\begin{align*}
f^{\Lam}= \Big(\sum_{k=1}^{\rho}\lambda_k f_{a_k}\Big)/\Big(\sum_{k=1}^{\rho}\lambda_k d_{a_k}\Big).
\end{align*}
Here, $\rho$, $\Omega_{\rm r}$, and $a_k$ are as defined in Definition \ref{def:Reduced_Domain}.
\end{defi}

Now, we are prepared to state the Cheeger problem equivalence requirements.
\begin{defi}\label{defi:unifiable}
As before, let $\digital=[d_k]_{1\leq k\leq m^2}$ be the measurements captured by sampling kernels $f_1,\dots,f_{m^2}$ with $0\leq d_i$, leading to $\rho$, $A$, and $\Omega_{\rm r}$ as in Definition \ref{def:Reduced_Domain}. For an arbitrary $\boldsymbol{\lambda}\in\Delta_{\rho}$, we define $I^{\Lam}=\alpha\chi_\shape$ to be the solution of \eqref{eq:CheegerConv} corresponding to the maximal Cheeger set $\shape$ with $f^{\Lam}$, when the domain is restricted to $\Omega_{\rm r}$. 
We call $(\digital;f_{1},\dots,f_{m^2})$
\emph{reducible} if $A$ can be partitioned into $K_1$ and $K_2$ such that
\begin{enumerate}[(i)]
\item $
\forall\,k\in K_1,\,\Lam\in\Delta_{\rho},\, \lambda_k=0:~~ \iint_{\Omega_{\rm r}} I^{\Lam}f_k\ud x\ud y<d_k,
$

\item
$
\forall\,k\in K_2,\,\Lam\in\Delta_{\rho}:~~ \iint_{\Omega_{\rm r}} I^{\Lam}f_k\ud x\ud y\leq d_k.
$

\end{enumerate}
\end{defi}

It was explained earlier that the measurements $d_i$ obtained from a binary shape through normalized sampling kernels satisfy $0\leq d_i \leq 1$. The requirements in Definition \ref{defi:unifiable} simply indicate that the maximal Cheeger solution corresponding to any convex combination of the kernels except a given one, should result in a strictly smaller measurement observed by the excluded kernel. Intuitively, we expect the Cheeger solution to have less contribution over the support of the excluded kernel. However, there are some exceptions; imagine the case where the support of a $3\times 3$ block of measurement kernels completely coincide with the interior of the binary shape. Thus, we shall have a block of all-one measurements. Now, it is likely that the maximal Cheeger set corresponding to a linear combination of the 8 surrounding kernels (but missing the middle one) using symmetric weights fully covers the support of the kernel in the middle. Hence, measuring this solution via the middle kernel results in $d_i=1$, instead of being strictly less than $1$. The partitions $K_1$ and $K_2$ in Definition \ref{defi:unifiable} are introduced to distinguish between the ordinary ($K_1$) and exceptional ($K_2$) cases. We postpone further discussion and clarifications about this definition to Section \ref{subsec:resolution}.



\begin{theo}\label{theo:ConsistentShape}
Let $(\digital;f_1,\dots,f_{m^2})$ be reducible according to Definition \ref{defi:unifiable}. Then, all solutions of the non-convex problem \eqref{eq:P_0} are included in the minimizers of its convex relaxation \eqref{eq:P_1}. Moreover, the solution set of \eqref{eq:P_1} contains a unique MCSMP.
\end{theo}

Our proof of Theorem \ref{theo:ConsistentShape} relies on the following lemma, the proof of which is provided in the appendix.


\begin{lem}\label{lem:mainLemma}
For a given dimension $n$ and a set $\{d_k\}_{k=1}^{n}\subset \R$, let $K_1,K_2$ be a partition of $\{1,\dots,n\}$, with the possibility of $K_1=\emptyset$ or $K_2=\emptyset$,  and let $v: \Delta_n \mapsto \R^n$ be a continuous function that satisfies
\begin{enumerate}[(i)]
\item \label{cod:linear_Comb}
$\forall\,\Lam\in\Delta_n:~~ \Lam^T\cdot v(\Lam) = \sum_{k=1}^{n}\lambda_k d_k,$

\item \label{cod:facets}
$\forall\,k\in K_1, \,\stackbin[\lambda_k=0]{}{\Lam\in\Delta_n}:~~ v_k(\Lam)<d_k$.

\item \label{cond:equality}
$\forall\, k\in K_2, \, \Lam\in\Delta_n:~~ v_k(\Lam)\leq d_k$
\end{enumerate}
Then, there exists $\Lam^*\in\Delta_n$ such that $v(\Lam^*)=[d_1,\dots,d_n]^T$.
\end{lem}


\noindent\textbf{Proof of Theorem \ref{theo:ConsistentShape}}.
The main ingredient of the proof is to show that under reducibility condition, \eqref{eq:P_0} and \eqref{eq:P_1} can be associated with a Cheeger problem. To show this, first note that $\consistency$ is essentially the same as $\mathcal{C}_{\Omega_{\rm r}}(\mathbf{D}_{\rm r};\,f_{a_1},\dots,f_{a_{\rho}})$, where $\rho,\, A,\, \Omega_{\rm r}$ are defined in Definition \ref{def:Reduced_Domain} and $\mathbf{D}_{\rm r}=[d_{k}]_{k\in A}$. In addition, for all $\Lam\in\Delta_{\rho}$, we have that
\begin{align*}
\mathcal{C}_{\Omega_{\rm r}}(\mathbf{D}_{\rm r};\,f_{a_1},\dots,f_{a_{\rho}}) \subseteq  \mathcal{C}_{\Omega_{\rm r}}(1;f^{\Lam})= BV_{\Omega_{\rm r}}(f^{\Lam}). 
\end{align*}
Therefore, any minimizer of \eqref{eq:CheegerConv} that falls inside $\mathcal{C}_{\Omega_{\rm r}}(\mathbf{D}_{\rm r};\,f_{a_1},\dots,f_{a_{\rho}})$ is also a minimizer of \eqref{eq:P_1}. Besides, if \eqref{eq:P_1} and \eqref{eq:CheegerConv} have a common minimizer, then, all the solutions of \eqref{eq:P_1} shall be among the solutions of \eqref{eq:CheegerConv}. This is indeed, what we aim to prove.

%

Let $I^{\Lam}$ be the maximal Cheeger set solution of \eqref{eq:CheegerConv} on $\Omega_{\rm r}$ corresponding to the weight kernel $f^{\Lam}$. Consider the function 
$$v(\Lam)\triangleq\Big[\iint_{\Omega_{\rm r}} I^{\Lam} f_{a_1}~,~\dots~,~\iint_{\Omega_{\rm r}} I^{\Lam} f_{a_{\rho}}\Big]^T.$$ 
We demonstrate that $v(\cdot)$ satisfies the conditions of Lemma \ref{lem:mainLemma}. The first condition directly follows from
$$1=\iint_{\Omega_{\rm r}} I^{\Lam} f^{\Lam}= \frac{1}{\sum_{k=1}^{\rho}\lambda_k d_{a_k}}\iint_{\Omega_{\rm r}} I^{\Lam} \sum_{k=1}^{\rho} \lambda_k f_{a_k} .$$ 
The reducibility property of $(\digital_{\rm r};f_{a_1},\dots,f_{a_{\rho}})$ also establishes Conditions (\ref{cod:facets}) and (\ref{cond:equality}) of Lemma \ref{lem:mainLemma}. Consequently, there exists $\Lam^*\in\Delta_{\rho}$ such that $$v_k(\Lam^*)=\iint_{\Omega_{\rm r}} I^{\Lam^*}f_{a_k}\ud x\ud y=d_{a_k},~~ 1\leq k\leq \rho.$$
%
This means that the bilevel maximal Cheeger solution $I^{\Lam^*}$, which minimizes \eqref{eq:CheegerConv}, is also consistent with the measurements $\digital_{A}$. Hence, $I^{\Lam^*}$ is also a minimizer of \eqref{eq:P_1} as well as \eqref{eq:P_0}; \emph{i.e.}, the three problems \eqref{eq:CheegerConv} with $f^{\Lam^*}$ over $\Omega_{\rm r}$, \eqref{eq:P_1} and \eqref{eq:P_0} share a minimizer. This proves the first part of the claim.

As for the second part, note that all minimizing shapes of \eqref{eq:P_0} are Cheeger solutions of \eqref{eq:CheegerConv}. Thus, their support should be included in the support of the maximal Cheeger solution $I^{\Lam^*}$. In words, $I^{\Lam^*}$ is a MCSMP.
%
%
$\hspace{\stretch{1}}\blacksquare$
\vspace{10pt}

Theorem \ref{theo:ConsistentShape} states that under reducibility, the solution set of \eqref{eq:P_1} is guaranteed to contain a MCSMP. Although we believe that the MCSMP is the unique solution of \eqref{eq:P_1} under reducibility, it is yet to be proven. However, we introduce a test in Theorem \ref{theorem:level-set} to verify whether an obtained solution to \eqref{eq:P_1} is the MCSMP. This test helps us in simulation results, where we implement a minimization technique and eventually obtain a solution with a numerical precision. First, it is difficult to make sure whether the result is precisely bilevel, and second, even if it is bilevel, is it the MCSMP? 

\begin{theo}\label{theorem:level-set}
Let $(\digital;f_1,\dots,f_{m^2})$ with $d_i=1$ for some $i$ be reducible (at least one measurement equal to one). If the point values of a solution to \eqref{eq:P_1} never exceed $1$, then, this solution is the MCSMP and it is binary (non-zero values are all $1$).
\end{theo}

\noindent\textbf{Proof.} 
Let $I(x,y)\leq 1$ be a solution to \eqref{eq:P_1}, and let  $i$ be the index of a measurement equal to $1$, \emph{i.e.}, $d_i=1$. By comparing $I(x,y)\leq 1$ and $d_i=1$, we conclude that for all $(x,y)\in\text{supp}(f_i)$ we should have $I(x,y)=1$ (the kernels are normalized). If $I$ is the MCSMP, as it takes the value $1$, it needs to be binary and the proof is complete. Therefore, let us assume the MCSMP to be $\tilde{I}\neq I$. As previously shown, the support of $\tilde{I}$ contains the support of $I$, which obviously contains the support of $f_i$. As $\tilde{I}$ is constant over its support and is also consistent with measurement $d_i$, we should have that $\tilde{I}(x,y)=1$ for all $(x,y)\in \text{supp}(f_i)$. Thus, $\tilde{I}$ is binary. However, this implies that $I$ never exceeds $\tilde{I}$ at any point, while they generate the same set of measurements. In turn, this suggests that $I$ cannot be less than $\tilde{I}$ on a set of non-zero measure. In other words, $I$ and $\tilde{I}$ are essentially equal at all points.
$\hspace{\stretch{1}}\blacksquare$

For recovering a binary shape from discrete measurements, we infer the following: when the sampling density is high enough to provide the reducibility condition for the measurements, the studied convex problem is potentially able to return a consistent binary shape with minimum perimeter. Besides, the boundary of the output shall be a $C^2$ curve.
 

\begin{figure}[]
\centering
\hfill
\subfloat[
]{\includegraphics[width=.35\linewidth]{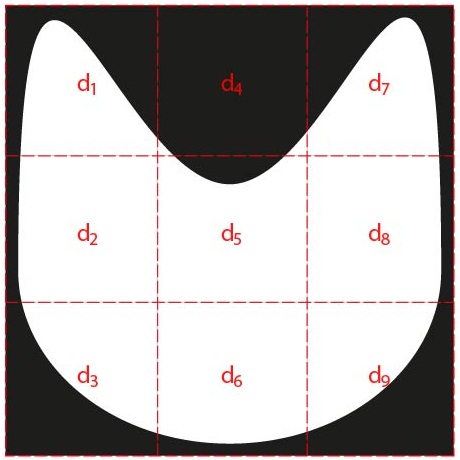}%
\label{fig:unifiablity_shape}}
\hfill
\subfloat[]{\includegraphics[width=.35\linewidth]{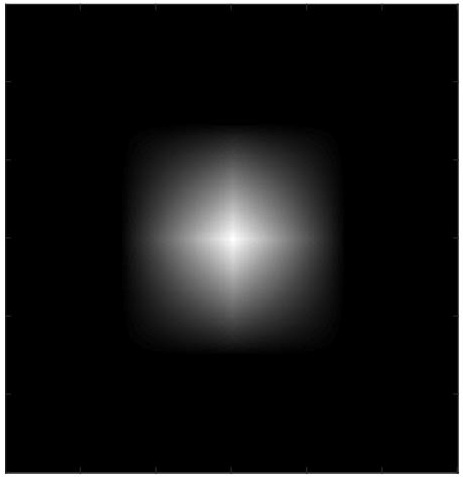}%
\label{fig:unifiablity_kernel}}
\hspace*{\fill}
\\
\hfill
\subfloat[
]{\includegraphics[width=.35\linewidth]{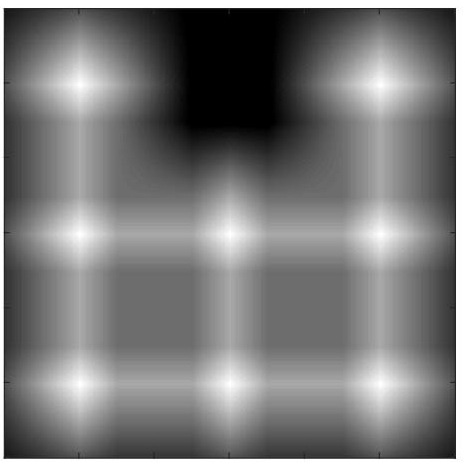}%
\label{fig:integrated_kernel}}
\hfill
\subfloat[
]{\includegraphics[width=.35\linewidth]
{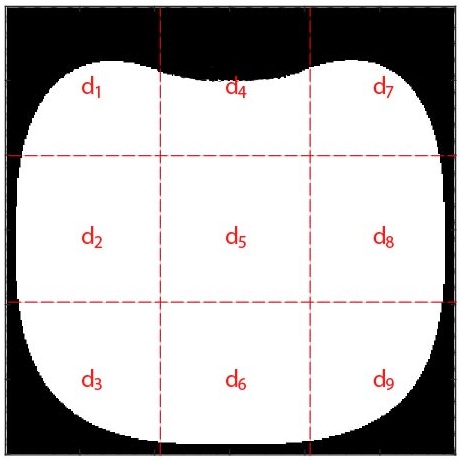}
\label{fig:unifiablity_Cheeger}}
\hspace*{\fill}
\caption{Violation of the reducibility criterion due to the low sampling density. (a) shows the original binary image over a $3\times 3$ sampling grid. This generates the measurement $d_4 = 0.0523$, when the sampling kernels are the shifts of the bilinear B-spline kernel in (b). (c) shows the reduced sampling kernel $f^{\Lam}$ corresponding to $\Lam =\lbrack 1,1,1,0,1,1,1,1,1\rbrack/8$, that results in the Cheeger solution (d) with levels $0$ and $0.9577$ (reproducing the larger measurement $d_4=0.5247$).
}
\label{fig:unifiable}
\end{figure}

\subsection{The Sampling Density Requirement}\label{subsec:resolution}
Earlier, we claimed that the reducibility condition in Definition \ref{defi:unifiable} is effectively a requirement on the minimum sampling density. Here, we illustrate this intuition by some examples.

First, we consider the sampling of the shape in Figure \ref{fig:unifiablity_shape} over a $3\times3$-pixel grid, employing the bilinear B-spline sampling kernel depicted in Figure \ref{fig:unifiablity_kernel}. 
This generates the measurements $$\digital = \begin{bmatrix}
0.5634 & 0.0523 & 0.5750\\
0.8996 & 0.9016 & 0.8882\\
0.5247 & 0.8817 & 0.5097
\end{bmatrix}.$$
Particularly, we focus on the $d_4$ measurement pixel (or $d_{12}$ in the usual matrix indexing format). It is evident that the value of this measurement is considerably lower than its neighboring measurement pixels. Intuitively, this sharp transition violates the resolution requirement. Now, we check the reducibility condition: let us exclude the $d_4$ pixel and apply equal weights for a convex combination of the remaining measurements, \emph{i.e.}, $\Lam =\lbrack 1,1,1,0,1,1,1,1,1\rbrack/8\in\Delta_9$. Figure \ref{fig:integrated_kernel} depicts the reduced sampling kernel $f^{\Lam}$, and Figure \ref{fig:unifiablity_Cheeger} shows the corresponding maximal Cheeger solution. Although $d_4$ did not contribute in this Cheeger solution, we observe substantial leakage over its region from the neighboring pixels. Thus, $(\digital,f_1,\dots,f_9)$ is not reducible. Oftentimes, sharp transitions between neighboring pixel values indicate lack of sufficient density for sampling the boundary curve of the shape (possibly, parts with high curvature). Similarly, the reducibility condition prevents the value of a pixel dropping substantially below its neighbors.


\begin{figure}[t]
\centering
\hfill
\subfloat[]{\includegraphics[width=.35\linewidth]{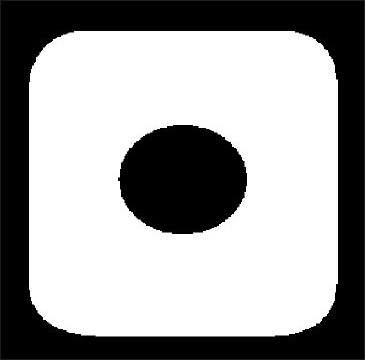}%
\label{fig:hole_shape}}
\hfill
\subfloat[]{\includegraphics[width=.35\linewidth]{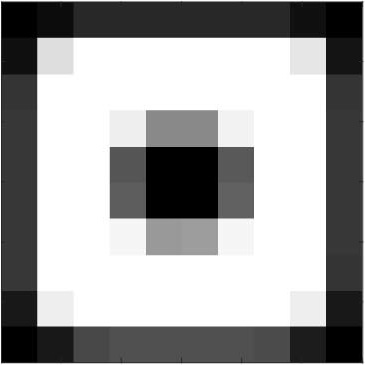}%
\label{fig:hole_discrete}}
\hspace*{\fill}
\\
\hfill
\subfloat[]{\includegraphics[width=.35\linewidth]{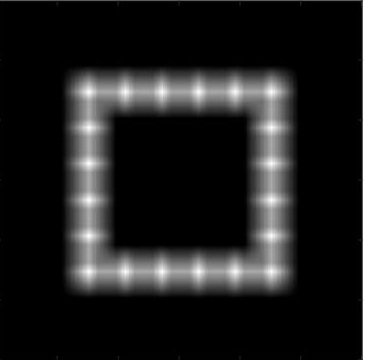}%
\label{fig:hole_kernel}}
\hfill
\subfloat[]{\includegraphics[width=.35\linewidth]{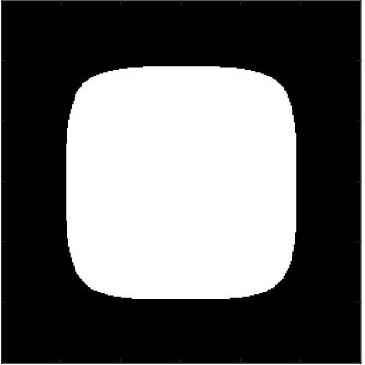}
\label{fig:hole_Cheeger}}
\hspace*{\fill}
\caption{The measurements of shapes with internal holes never satisfy the reducibility requirement, no matter how high is the measurement density, unless the original domain is replaced with the reduced domain. (a) A binary shape image with an internal hole, (b) the corresponding $10\times10$ measurements, with $d_{55}=d_{56}=d_{65}=d_{66}=0$, (c) reduced kernel $f^{\Lam}$, with equal contributions from the 20 kernels associated with pixels on the borders of the central $6\times6$ sub-grid, and (d) the maximal Cheeger solution with levels 0 and 0.9891.}
\label{fig:hole}
\end{figure}

One of the shortcomings of reformulating \eqref{eq:P_1} as \eqref{eq:CheegerConv} using a single reduced kernel $f^{\Lam}$ is that the Cheeger solution never admits a hole. Figure \ref{fig:hole} provides a pictorial explanation. Here, we would like to recover the shape image in Figure \ref{fig:hole_shape} from its discrete measurements on a $10\times 10 $ grid. The hole causes four vanishing middle pixels (Figure \ref{fig:hole_discrete}), which make it obvious that the shape content is $0$ in the middle. We now consider a reduced kernel by linearly combining (with equal weights) only the $20$ kernels associated with the pixels on the perimeter of the central $6\times 6$ sub-grid (Figure \ref{fig:hole_kernel}). As claimed, the Cheeger solution to \eqref{eq:CheegerConv} depicted in Figure \ref{fig:hole_Cheeger} has no holes and completely covers the middle part. This seems to violate the reducibility condition, no matter how high we set the sampling density. However, note that we remove the $0$ pixels from the domain in Definition \ref{defi:unifiable}. Therefore, the Cheeger solution over the reduced domain is forced to have a hole, although it is not considered as hole with respect to the reduced domain.

\begin{figure*}[t]
\centering
\hfill
\subfloat[]{\includegraphics[width=.3\linewidth]{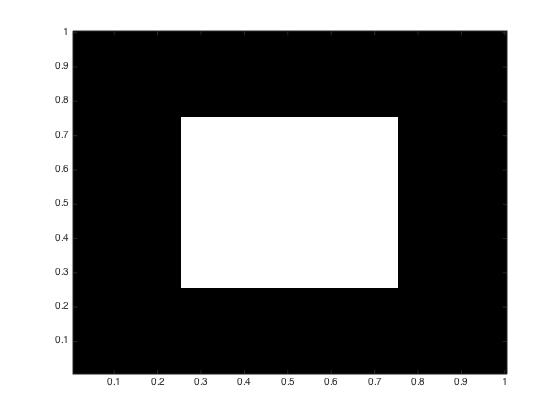}%
\label{fig:boxSpline}}
\hfill
\subfloat[]{\includegraphics[width=.3\linewidth]{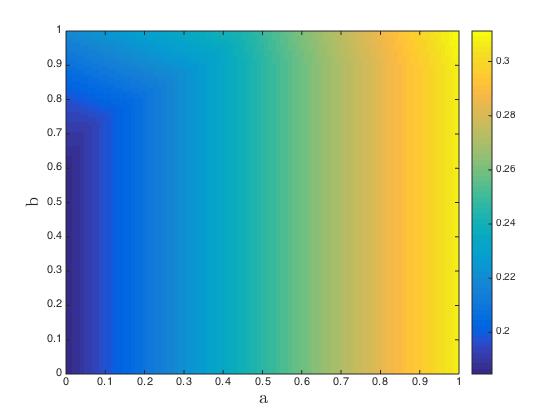}%
\label{fig:}}
\hfill
\subfloat[]{\includegraphics[width=.3\linewidth]{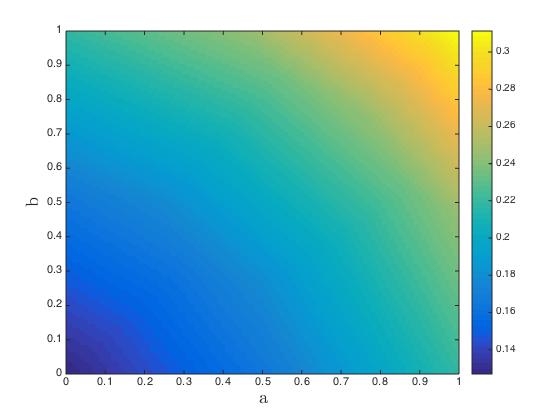}
\label{fig:}}
\hspace*{\fill}
\caption{The approximate functions (b) $Y$ and (c) $Z$ for the characterization of reducible measurements $\digital_{2\times2}$ when the kernels $f_1,f_2,f_3,f_4$ are shifts of the box-spline kernel in (a) centered at point $(0.25,0.75),(0.25,0.25),(0.75,0.75)$ and $(0.75,0.25)$, respectively.}
\label{fig:unifiablity_boxSpline}
\end{figure*}

\begin{figure*}[t]
\centering
\hfill
\subfloat[]{\includegraphics[width=.3\linewidth]{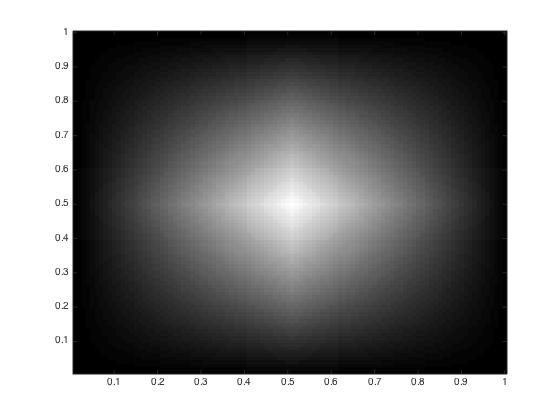}%
\label{fig:bilinearSpline}}
\hfill
\subfloat[]{\includegraphics[width=.3\linewidth]{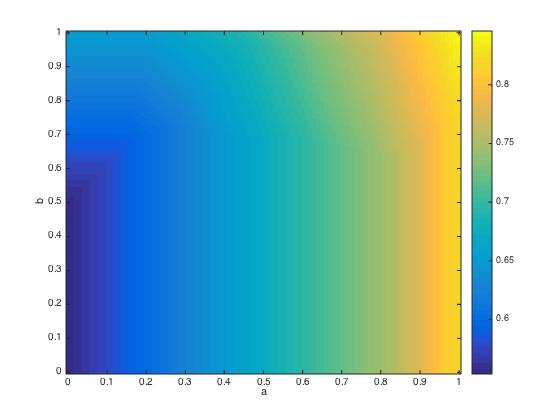}%
\label{fig:}}
\hfill
\subfloat[]{\includegraphics[width=.3\linewidth]{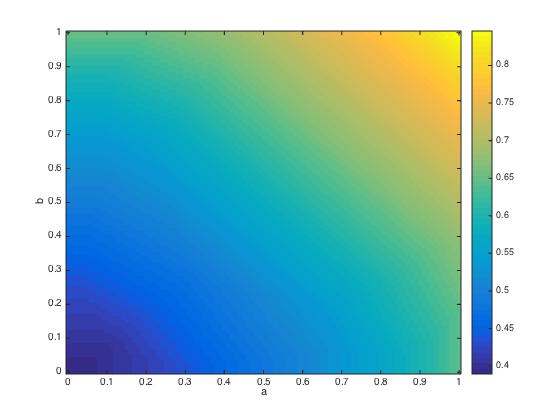}%
\label{fig:}}
\hspace*{\fill}
\caption{The approximate functions (b) $Y$ and (c) $Z$ for the characterization of reducible measurements $\digital_{2\times2}$ when the kernels $f_1,f_2,f_3,f_4$ are shifts of the bilinear B-spline kernel in (a) centered at point $(0.25,0.75),(0.25,0.25),(0.75,0.75)$ and $(0.75,0.25)$, respectively.}
\label{fig:unifiablity_bilinearSpline}
\end{figure*}

The reducibility condition in Definition \ref{defi:unifiable} is a useful guarantee for recovering a shape image. However, verifying it for a given set of measurements and sampling kernels is a combinatorial problem in general. For the purpose of illustration, we investigate the simple case with $2\times 2$ measurement pixels. Let $\digital=\begin{bmatrix} d_1&d_3\\d_2&d_4\end{bmatrix}$ with elements in $[0,1]$ represent the measurement matrix. Without loss of generality, we assume that $d_4=\rho\leq 1$ is the largest element. To verify the reducibility condition, we need to exclude each pixel, apply an arbitrary convex combination on the rest and check an inequality. As we can categorize $d_4$ to the $K_2$ set in Definition \ref{defi:unifiable}, the inequalities when $d_4$ is excluded are trivial. To verify other inequalities, note that we can scale all measurements by the factor $\tfrac{1}{\rho}$ (or any other positive real). In fact, the scaling does not affect the support set of the Cheeger solutions. Consequently, the reducibility condition for $\digital$, boils down to a set of inequalities on each of $\frac{d_1}{\rho},\frac{d_2}{\rho},\frac{d_3}{\rho}$ in terms of the other two:
\begin{align*}
d_1 & > \rho Z(\frac{d_2}{\rho},\frac{d_3}{\rho}),\\
d_2 & > \rho Y(\frac{d_1}{\rho},\frac{d_3}{\rho}),\\
d_3 & > \rho Y(\frac{d_1}{\rho},\frac{d_2}{\rho}).
\end{align*}
The symmetries of the problem indicate that the lower-bounds on $d_2$ and $d_3$ can be represented using the same function ($Y(\cdot,\cdot)$), and the lower-bound $Z(\cdot,\cdot)$ on $d_1$ is symmetric with respect to the two inputs. In Figures  \ref{fig:unifiablity_boxSpline} and \ref{fig:unifiablity_bilinearSpline} we depict the functions $Y,Z$ for two choices of the sampling kernel, namely, the box-spline (Figure \ref{fig:unifiablity_boxSpline}) with non-overlapping kernels and bilinear B-spline kernels with $50\%$ overlap (Figure \ref{fig:unifiablity_bilinearSpline}). 
The overlap introduces correlation among the neighboring pixels, which naturally leads to tighter regions for validity of the reducibility condition. This is indicated by larger $Y$ and $Z$ values. For instance, the measurement set $\digital=\begin{bmatrix}0.576 & 0.72\\ 0.216&0.216 \end{bmatrix}$ is reducible under the box-spline sampling kernels, but not under the bilinear B-spline kernels.

\section{Numerical Experiments}\label{sec:implementation}
In this section, we aim at numerically calculating the optimal solution(s) of the convex problem \eqref{eq:P_1}. For this purpose, we restrict the simulations to the discrete setting. Below, we first explain the equivalent problem in the discrete domain and then, present the simulation results.

\subsection{Discrete Formulation}
For conducting computer simulations, we are limited to discrete scenarios. Therefore, we discretize the domain $\domain$ (and subsequently all the functions defined on $\domain$) with a finite step-size $h\sim \frac{1}{N}$ for some large integer $N$. This will approximate $\domain$ and the continuous-domain objects $I,f_1,...,f_{m^2}$ by their pseudo samples at the 2D grid
\begin{align*}
\{(ih,jh);~i,j=1,2,...,N\}
\end{align*}
resulting in $\mathbb{R}^{N\times N}$ matrices. In the discretized version, we approximate the gradient operator by evaluating the forward differences; for instance we approximate $\nabla I$ with an $\mathbb R^{N\times N\times 2}$ tensor defined as
\begin{align}
(\nabla I)_{i,j,k}=(\nabla I)^k_{i,j}\label{eq:nabla1}
\end{align}
where
\begin{align}
&(\nabla I)^1_{i,j}=\left\{\begin{array}{ll}
I_{i+1,j}-I_{i,j}&\text{if }i<N,\\
0&\text{if }i=N,\end{array}\right.\label{eq:nabla2}\\
&(\nabla I)^2_{i,j}=\left\{\begin{array}{ll}
I_{i,j+1}-I_{i,j}&\text{if }j<N,\\
0&\text{if }j=N.\end{array}\right.\label{eq:nabla3}
\end{align}
It is shown that in the asymptotic regime of $N\to\infty$, the results obtained with the discretized model converge to their continuous-domain counterpart introduced in \eqref{eq:P_1} \cite{Chambolle2004}. 

\begin{figure*}[htb!]
\centering
\subfloat[]{\includegraphics[width=0.23\linewidth]{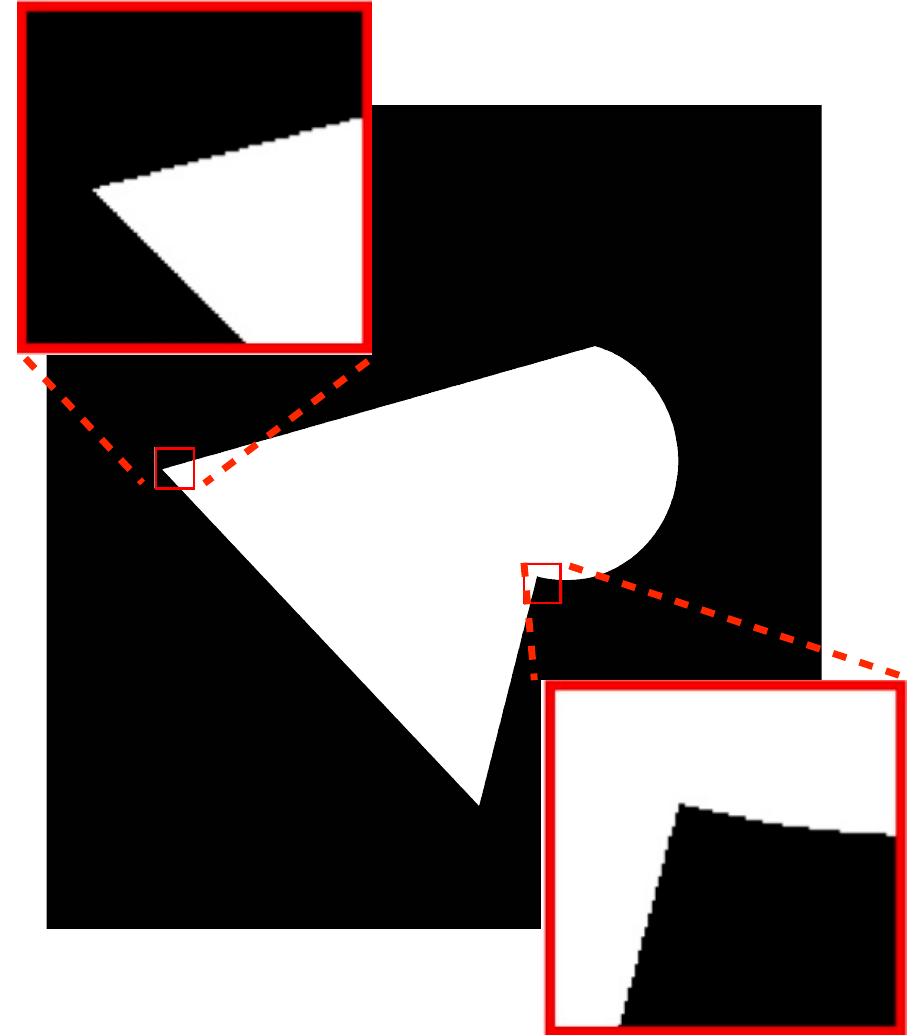}\label{fig:object_original}}
\hfill
\subfloat[]{\includegraphics[width=0.23\linewidth]{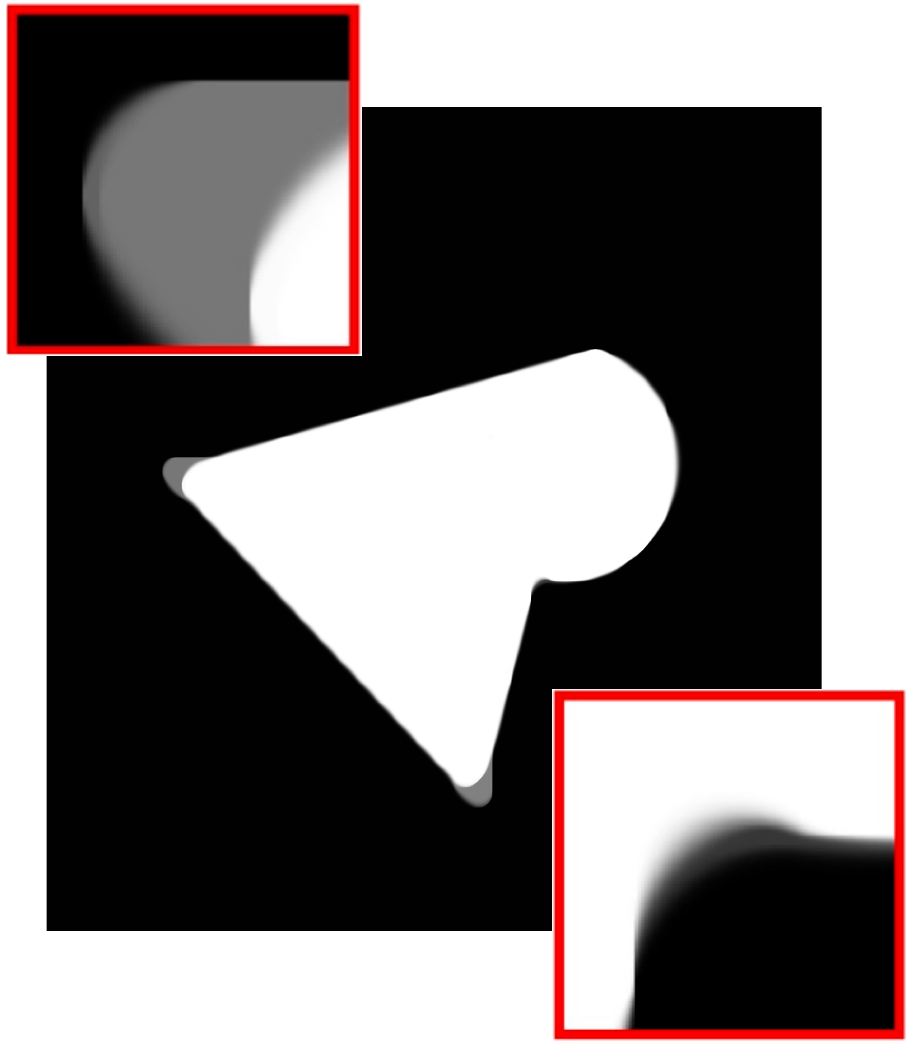}\label{fig:object_hat40}}
\hfill
\subfloat[]{\includegraphics[width=0.23\linewidth]{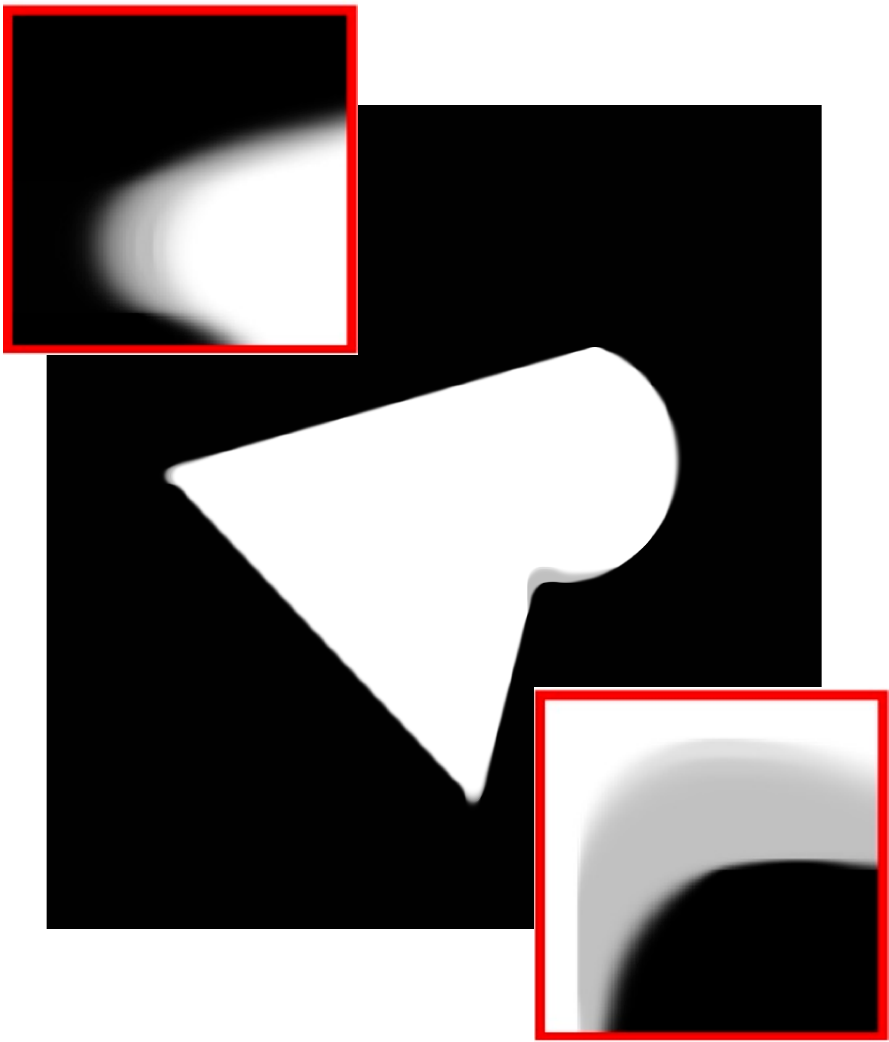}\label{fig:object_hat50}}
\hfill
\subfloat[]{\includegraphics[width=0.23\linewidth]{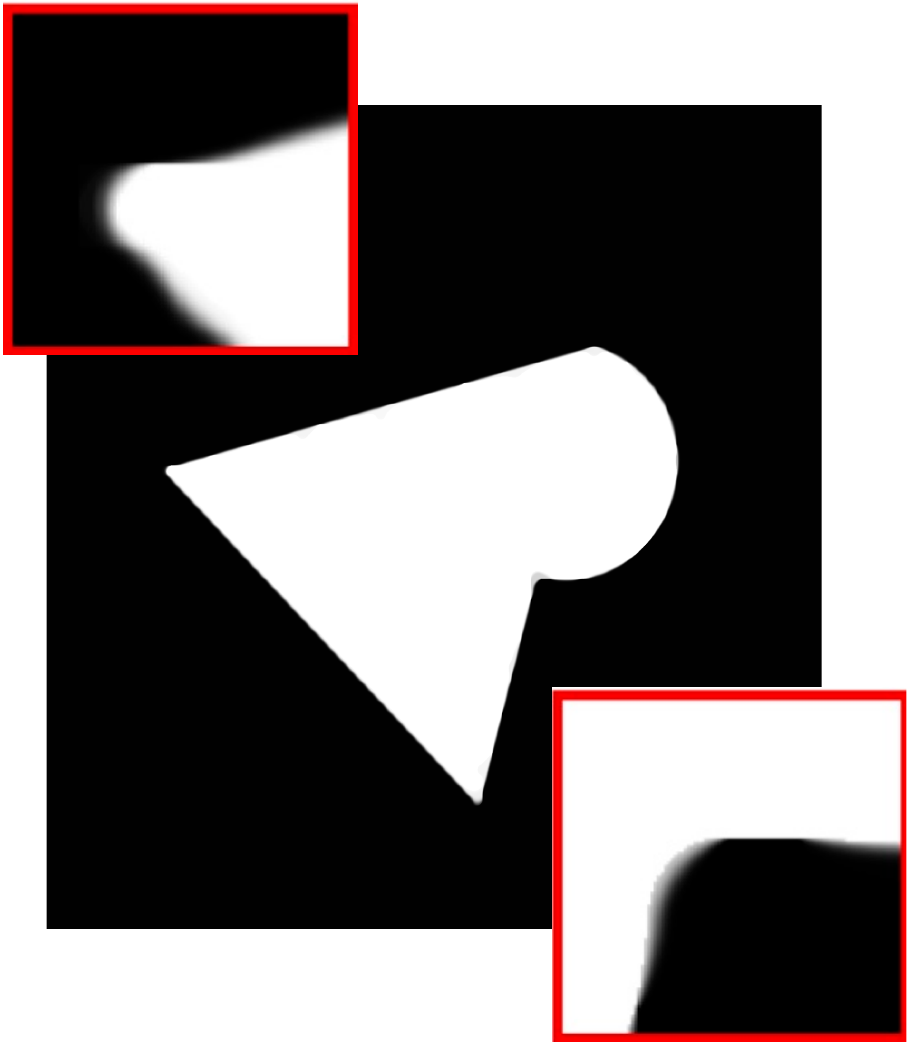}\label{fig:object_hat80}}
\caption{The performance of algorithm \eqref{eq:P_1} in shape recovery: (a), (b), (c) and (d) show the original shape image displayed with the resolution $2000\times 2000$ and its approximations using $40\times 40$, $50\times 50$ and $80\times 80$  measurements, respectively. Note that the reconstructed images are binary only when the number of measurements is large enough.}
\label{fig:analytical_object}
\end{figure*}

One of the standard approaches for solving the associated discrete minimization is the gradient descent algorithm, which is rather slow in high dimensions (small step-size $h$). In this paper, we use the recently proposed primal-dual algorithm in \cite{Strekalovskiy2014,Pock2009}, which is significantly faster and enjoys convergence guarantees. This scheme is based on the weak formulation of the total variation:
\begin{align}
\min_{I\in\mathcal C}\iint g~|\nabla I|&=\min_{I\in\mathcal C}~\max_{|\zeta|_2\leq g}\Big\{\iint -I~\text{div}\zeta\Big\}\nonumber\\
&=\min_{I\in\mathcal C}~\max_{|\zeta|_2\leq g}\langle-I,\text{div}\zeta\rangle
\end{align}
with the dual variable $\zeta: \Omega\to\mathbb{R}^2$. Here, $\text{div}\zeta$ stands for the divergence and is defined as the negative of the gradient adjoint. Each iteration of the optimization algorithm alternates between a gradient descent and a gradient ascent on the primal and dual variables, respectively. In short, the update equations are as follows:
\begin{align}
&\zeta^{(k+1)}=\text{Proj}_{\mathcal B(g)}(\zeta^k+\sigma_k\nabla\overline{I}^{(k)}),\label{eq:step1}\\
&I^{(k+1)}=\text{Proj}_{\mathcal C_\Omega(\digital;f_1,...,f_{m^2})}(I^{(k)}+\tau_k\text{div}\zeta^{(k+1)}),\label{eq:step2}\\
&\theta_k=\frac{1}{\sqrt{1+4\tau_k}},~\tau_{k+1}=\theta_k\tau_k,~\sigma_{k+1}=\sigma_k/\theta_k,\\
&\overline{I}^{(k+1)}={I}^{(k+1)}+\theta_k({I}^{(k+1)}-{I}^{(k)}),
\end{align}
where $k$ represents the iteration index.
Here, the notation $\text{Proj}_A(\cdot)$ stands for the orthogonal projection of the argument onto the set $A$ and $\mathcal B(g)$ represents the ball with radius $g$ in the space of $N\times N\times 2$ tensors:
\begin{align*}
\mathcal B(g)=\Big\{u\in\mathbb R^{N\times N\times 2}~; \sqrt{u_{i,j,1}^2+u_{i,j,2}^2}\leq g_{i,j}\Big\}.
\end{align*} 
Hence, $\text{Proj}_{\mathcal B(g)}(\cdot)$  scales only the points outside the ball $\mathcal B(g)$. The somewhat more complicated projection of  $\text{Proj}_{\mathcal C_\Omega(\digital;f_1,...,f_{m^2})}(\cdot)$ is also implemented using the POCS algorithm \cite{Bregman67}.
The initial values $I^{(0)}$ and $\zeta^{(0)}$ are arbitrary, with $\overline{I}^{0}=I^{(0)}$ and time steps $\tau_0\sigma_0\|\nabla\|^2<1$ \cite{Strekalovskiy2014}. By analogy (continuous setting), the divergence in \eqref{eq:step2} shall be the negated adjoint of the discrete gradient used in \eqref{eq:step1}. For the forward difference gradient in equations \eqref{eq:nabla1}-\eqref{eq:nabla3}, this leads to 
\begin{align*}
(\text{div}\zeta)_{ij}=&\left\{\begin{array}{ll}
\zeta^1_{i,j}-\zeta_{i-1,j}^1&\text{if }1<i<N,\\
\zeta_{i,j}^1&\text{if }i=1,\\
-\zeta^1_{i-1,j}&\text{if }i=N,\end{array}\right.\\
+&\left\{\begin{array}{ll}
\zeta^2_{i,j}-\zeta_{i,j-1}^2&\text{if }1<j<N,\\
\zeta_{i,j}^2&\text{if }j=1,\\
-\zeta^2_{i,j-1}&\text{if }j=N.\end{array}\right.
\end{align*}

\subsection{Simulation Results}
In the first experiment, we study the effect of the number of measurements on the reconstructed images obtained with the proposed algorithm. Recalling the result of the previous section, we expect the solution of \eqref{eq:P_1} to be binary, given adequate number of measurement pixels. In this experiment, we employ a shape image with a parametric description, composed of a semicircle laid on one side of an equilateral triangle (Figure \ref{fig:object_original}). This enables us to precisely access and display the image at arbitrary fine resolutions as a reference. Figure \ref{fig:object_original} shows the image at the resolution $2000\times 2000$. Figs. \ref{fig:object_hat40}, \ref{fig:object_hat50} and \ref{fig:object_hat80} show the solutions of algorithm \eqref{eq:P_1} with the same resolution applied to the measurement of sizes $40\times 40$, $50\times 50$ and $80\times 80$, respectively. All measurements are generated with a box-spline kernel. The original shape has non-smooth details around the corners and thus, to facilitate comparison, we enlarged the reconstructed images around these areas. The results reveal that with lack of enough measurements, the reconstructed images have more than two levels. It seems that the $80\times 80$-pixel image provides enough measurements to have a binary optimal solution for \eqref{eq:P_1}.  

\begin{figure}[htbp!]
\centering
\includegraphics[width=0.8\linewidth]{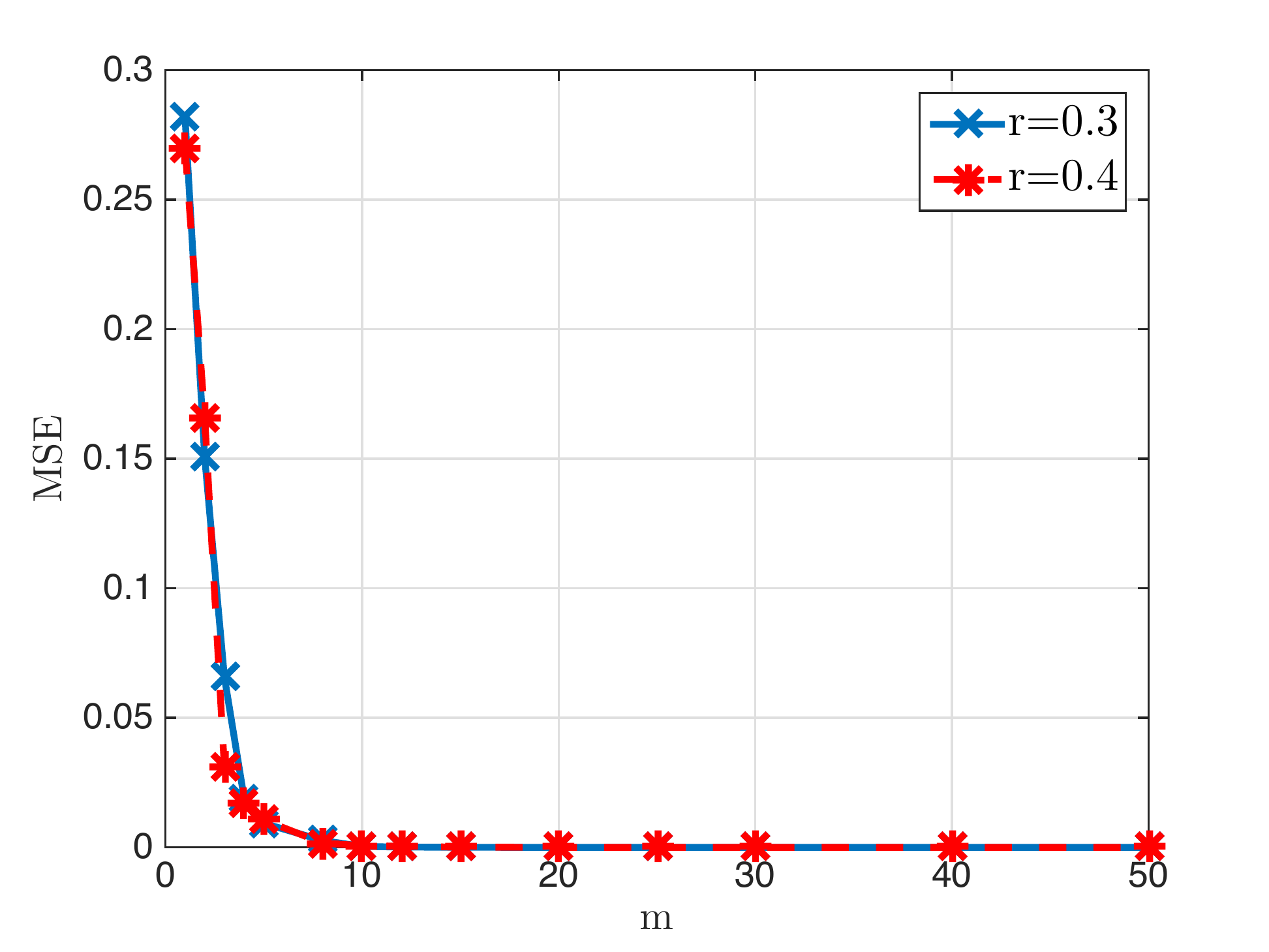}
\caption{The average MSE of the recovered binary images constrained by $m\times m$ samples of random circles with fixed radius for two radii $r=0.3$ and $r=0.4$.}
\label{fig:plot}
\end{figure}

In a similar experiment, we examine the performance of our algorithm in recovering circles from different number of measurements. For this purpose, we run a Monte Carlo experiment by generating 20 circles with fixed radius and random centers in the image plane. We then consider outputs of the algorithm at resolution $600\times 600$ constrained with $m\times m$ analytic measurements of the circles with box-spline PSFs and different values of $m$. Figure \ref{fig:plot} shows the average mean squared errors of the reconstructed images (after thresholding at level 0.5) versus $m$ for two different radii. The plots in this figure clearly indicate that the algorithm always perfectly recovers the circles from $m\times m$ measurements when $m$ is greater than 10. 
\begin{figure*}[htb!]
\centering
\hfill
\begin{minipage}[t]{0.45\linewidth}
\subfloat[]{\includegraphics[height=0.45\linewidth,valign=b]{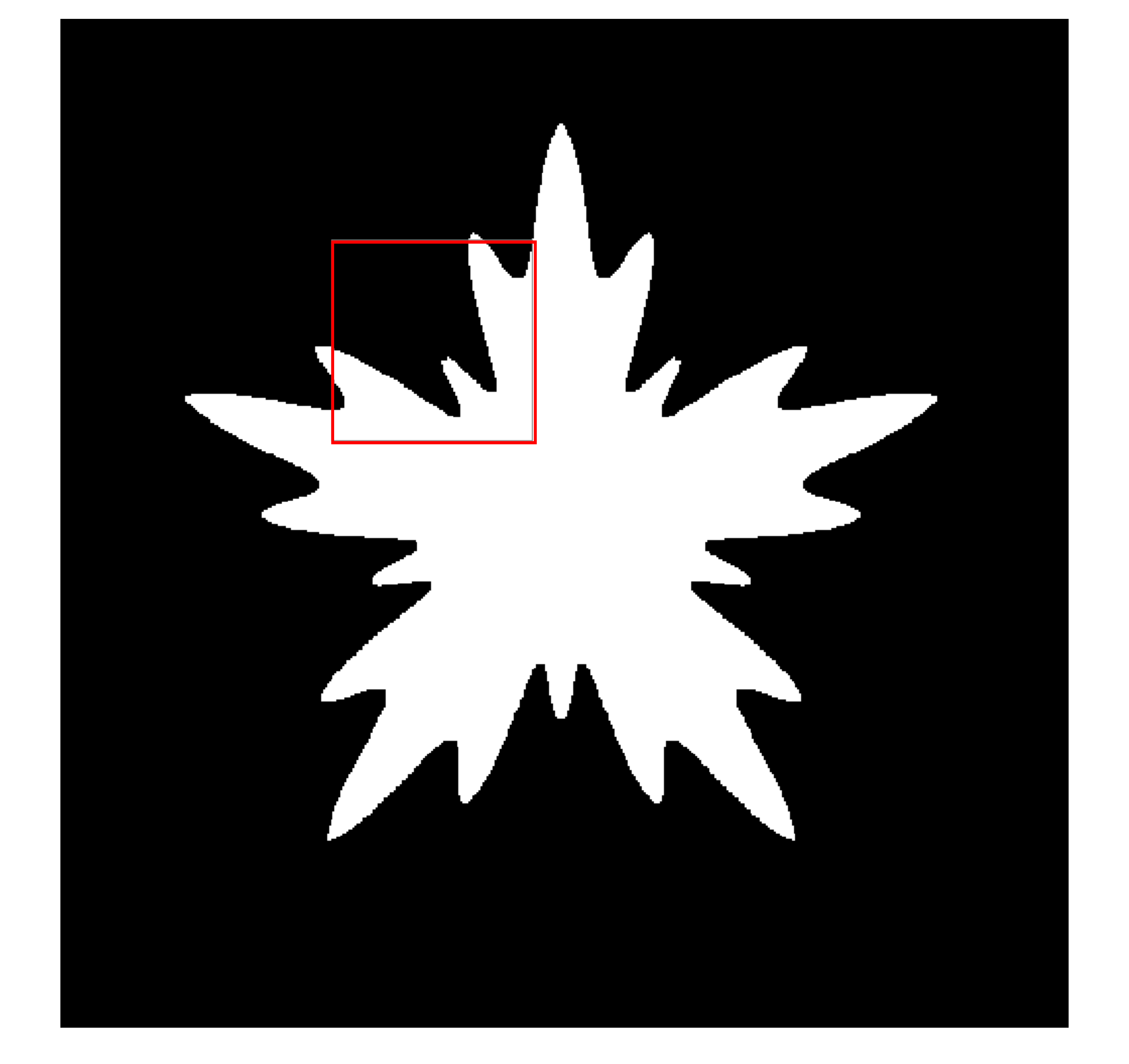}\label{fig:device}}
\subfloat[]{\includegraphics[height=0.39\linewidth,valign=b]{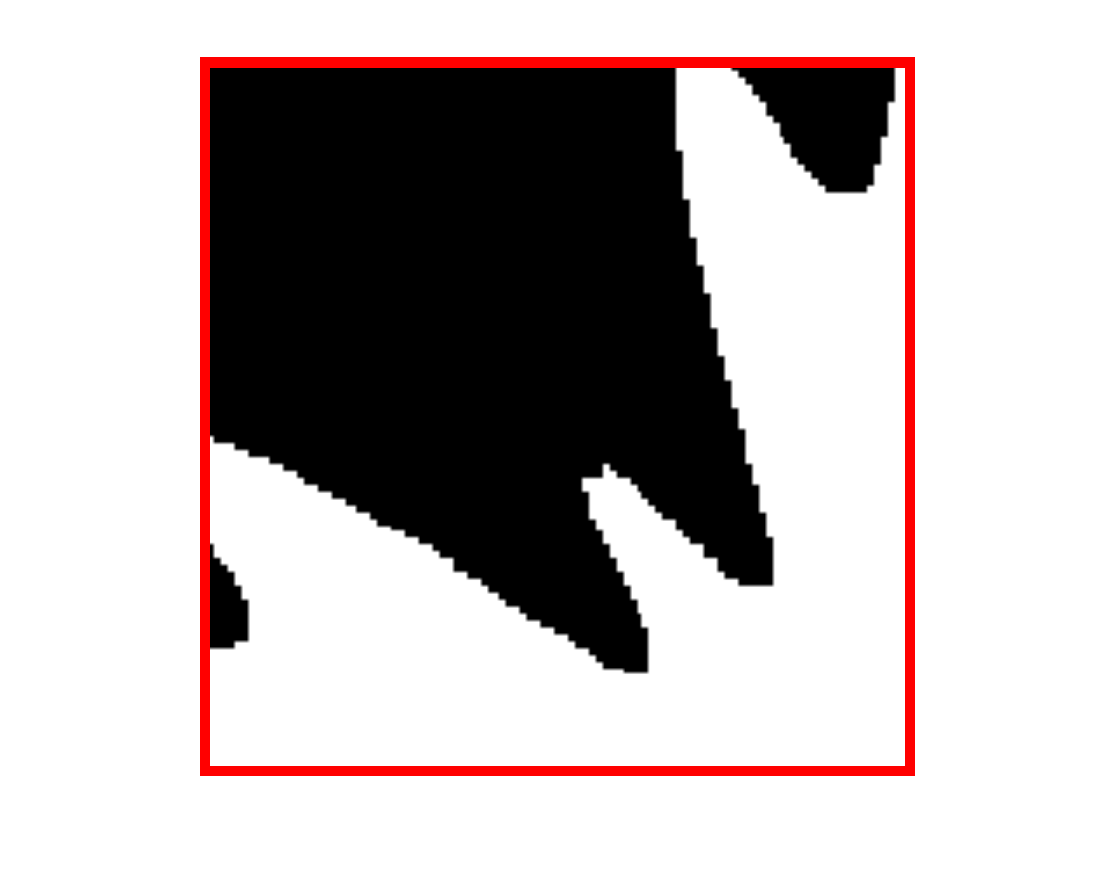}\label{fig:device_zoom}}
\end{minipage}
\hfill
\begin{minipage}[t]{0.45\linewidth}
\subfloat[]{\includegraphics[height=0.45\linewidth,valign=b]{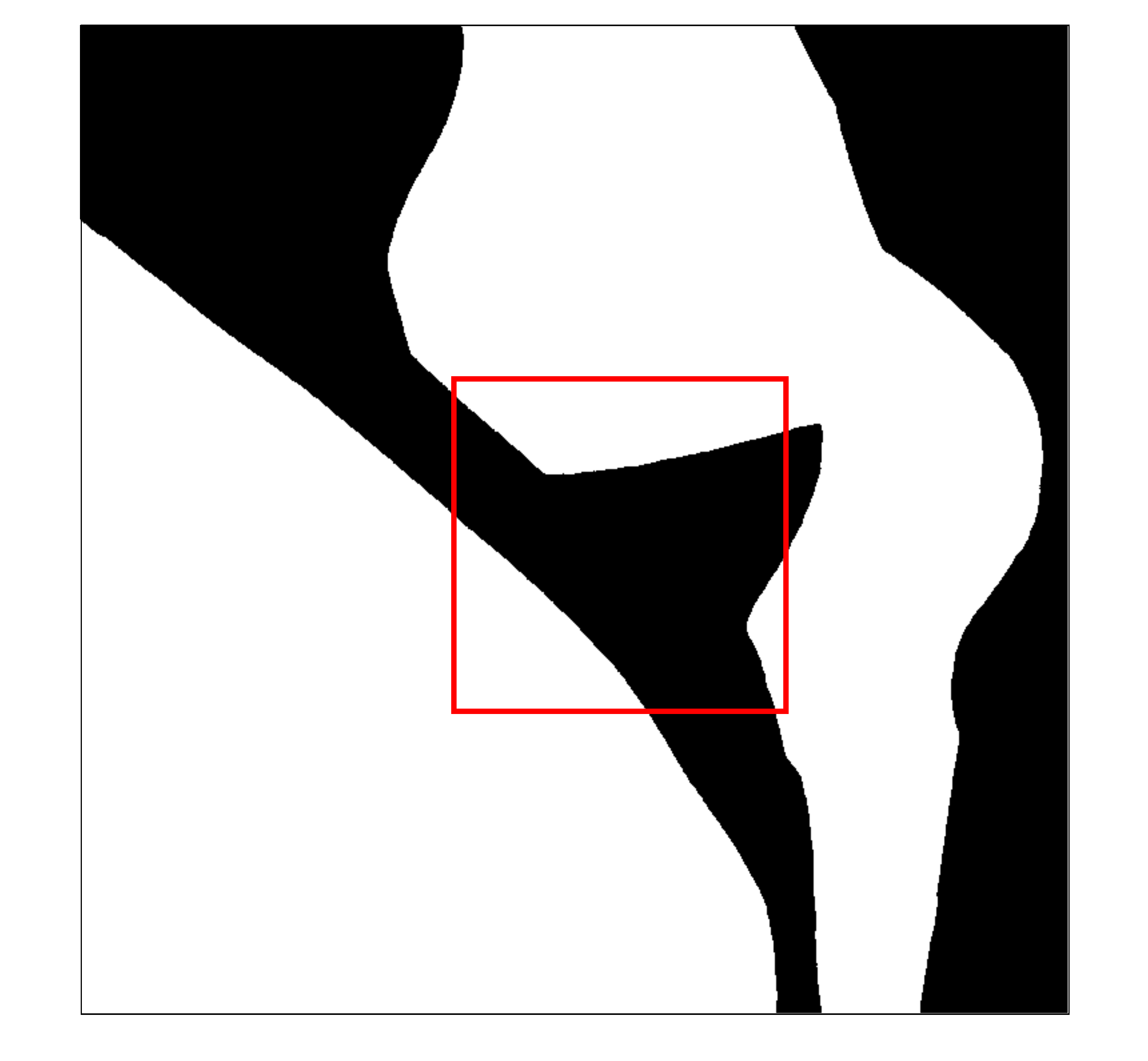}\label{fig:Lady}}
\subfloat[]{\includegraphics[height=0.39\linewidth,valign=b]{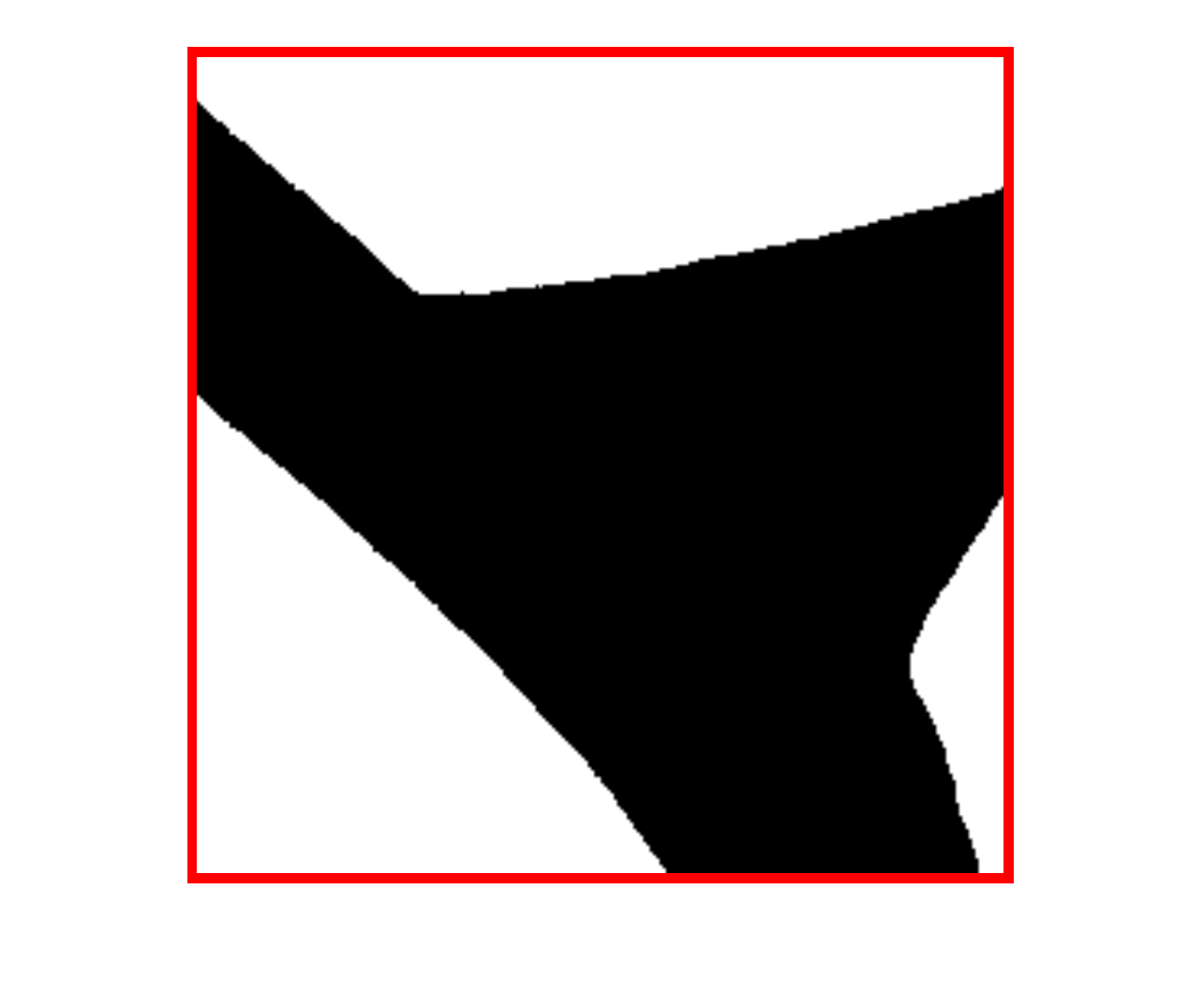}\label{fig:Lady_zoom}}
\end{minipage}
\hfill
\\
\hfill
\begin{minipage}[t]{0.45\linewidth}
\subfloat[]{\includegraphics[height=0.39\linewidth]{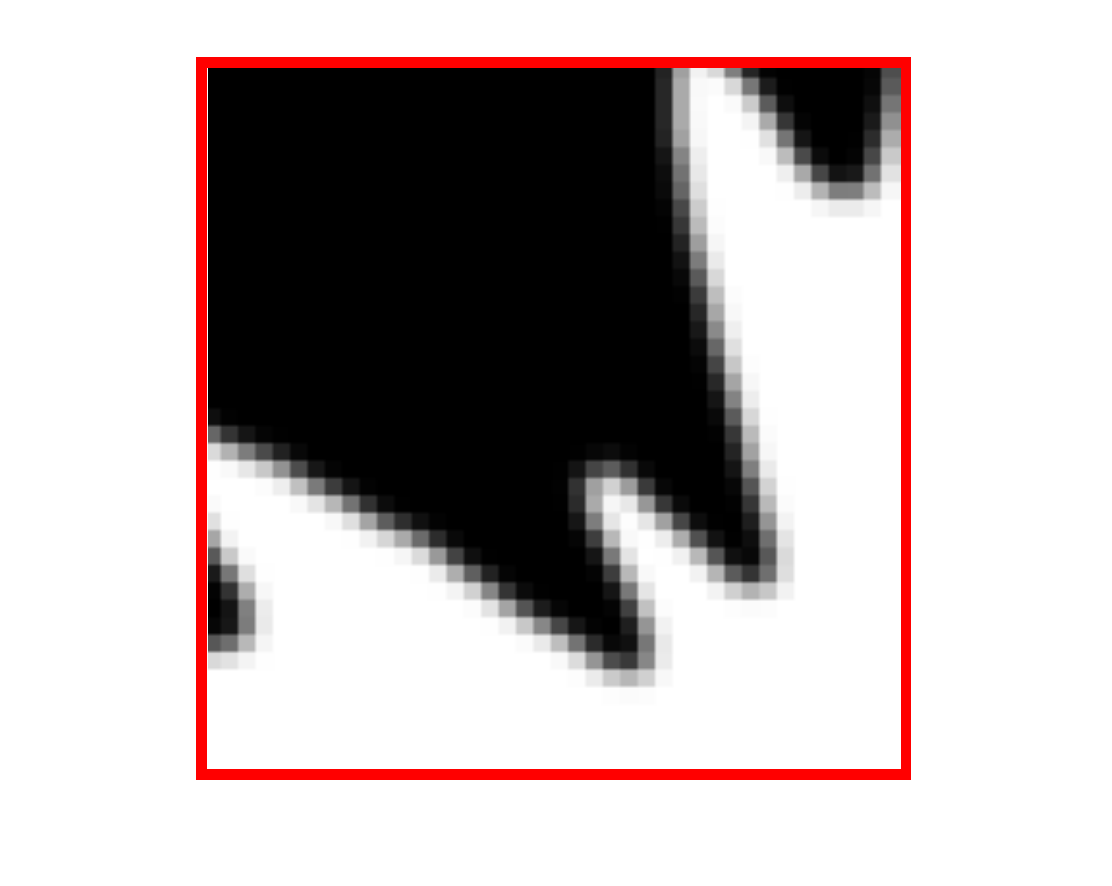}\label{fig:device_image_zoom}}
\subfloat[]{\includegraphics[height=0.39\linewidth]{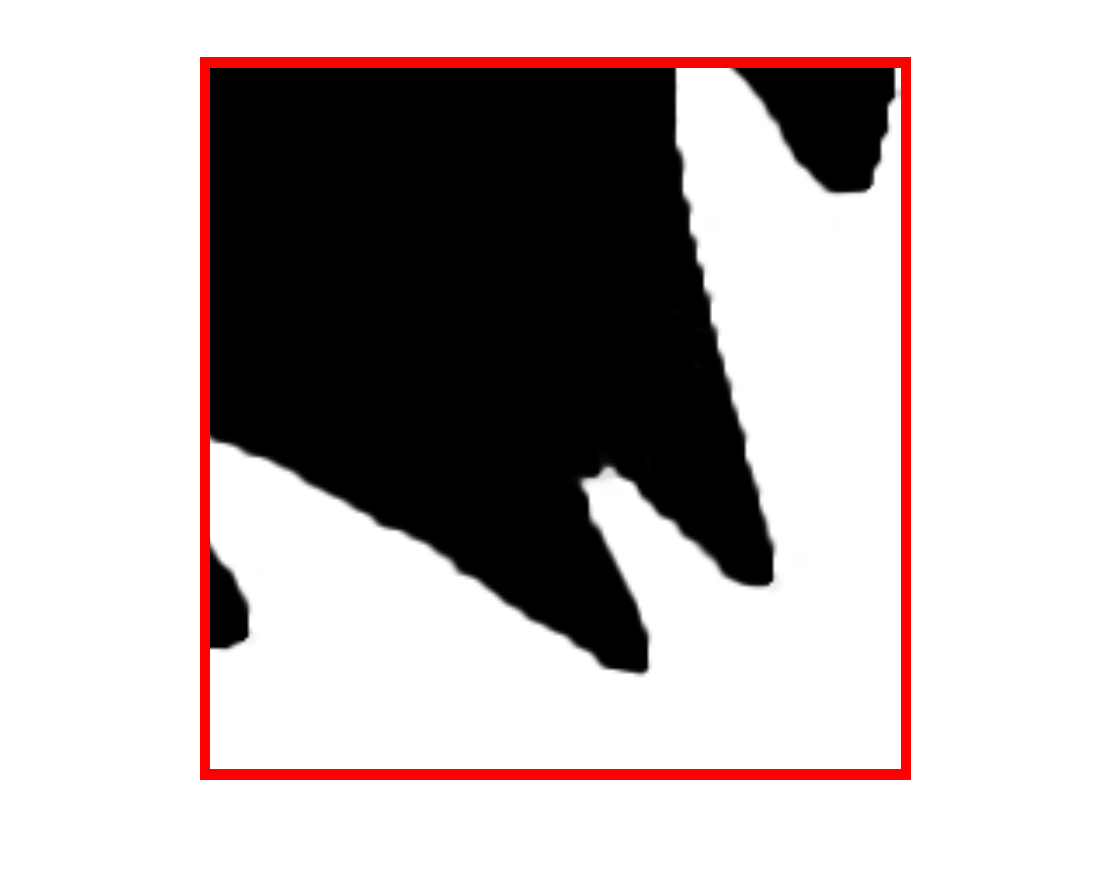}\label{fig:device_hat_zoom}}
\end{minipage}
\hfill
\begin{minipage}[t]{0.45\linewidth}
\subfloat[]{\includegraphics[height=0.39\linewidth,valign=b]{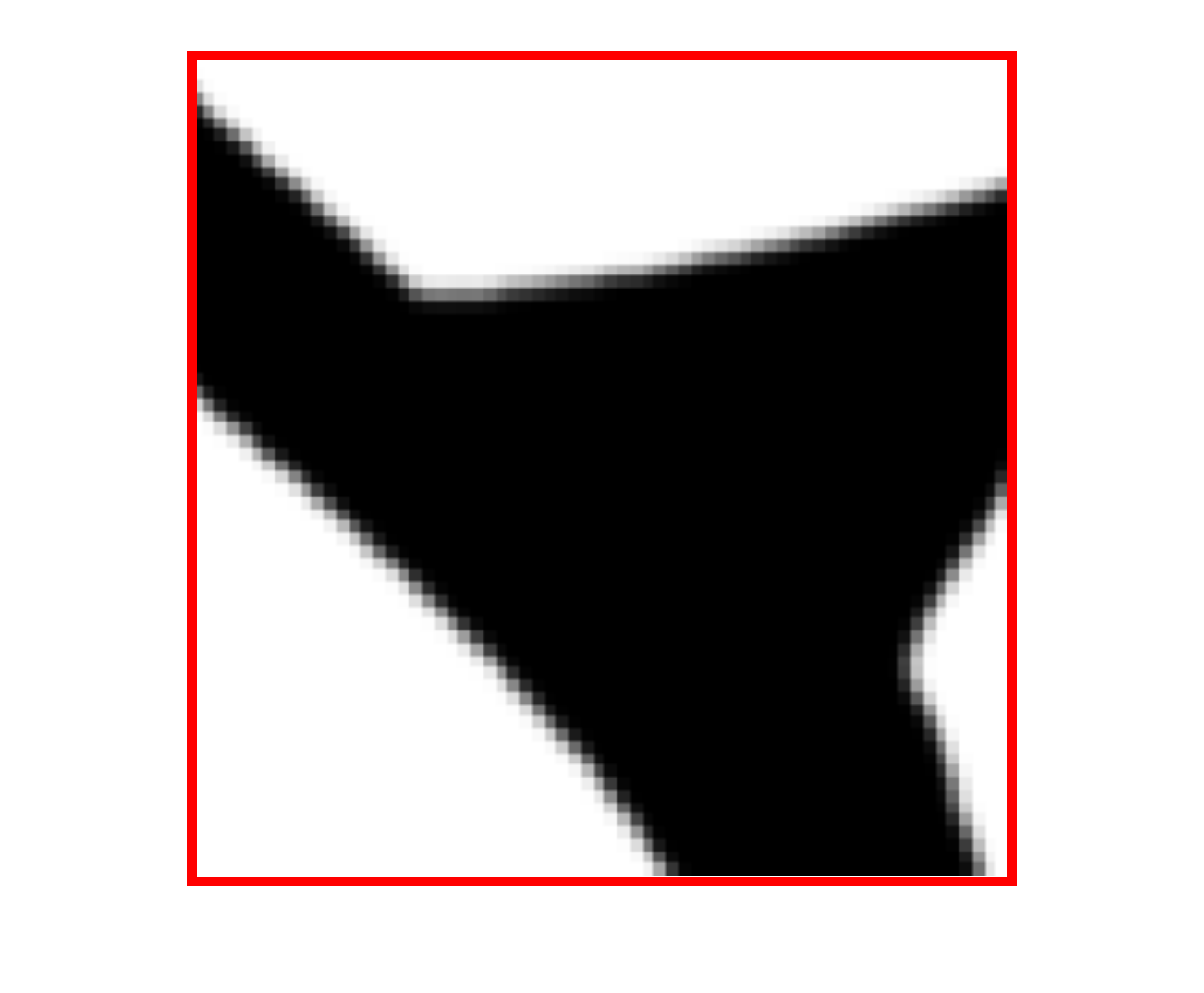}\label{fig:Lady_image_zoom}}
\subfloat[]{\includegraphics[height=0.39\linewidth,valign=b]{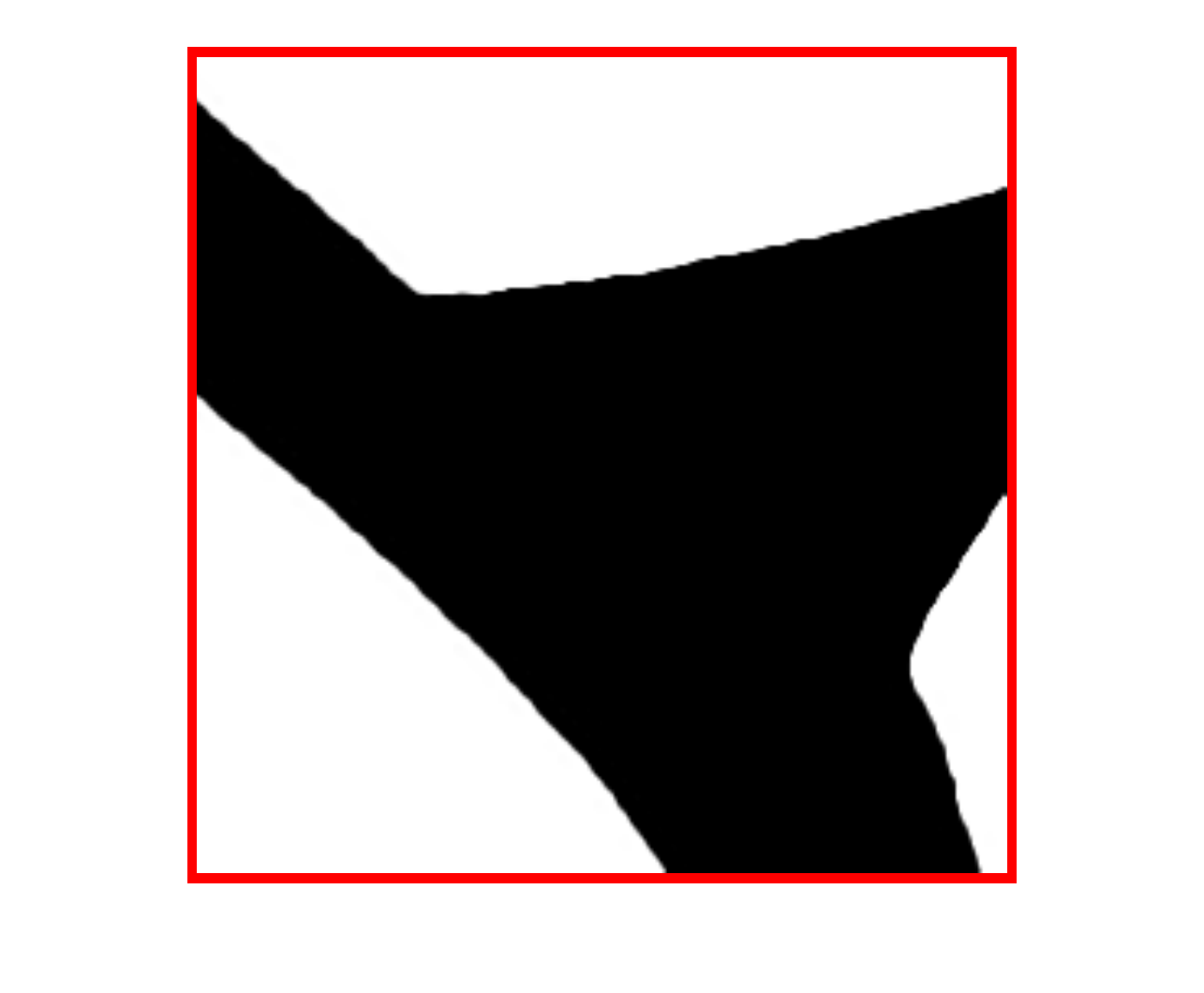}\label{fig:Lady_hat_zoom}}
\end{minipage}
\hfill
\caption{Consistent shape reconstruction with the proposed algorithm: Figs. \ref{fig:device} and \ref{fig:Lady} display two shapes at the resolution $1000\times 1000$ that will be approximated from $200\times 200$ discrete images, generated with biquadratic B-spline sampling kernels. Figs. \ref{fig:device_zoom}, \ref{fig:device_image_zoom} and \ref{fig:device_hat_zoom} show the same enlarged sections of the original shape \ref{fig:device}, the corresponding discrete image and reconstructed shape, respectively.  Figs. \ref{fig:Lady_zoom}, \ref{fig:Lady_image_zoom} and \ref{fig:Lady_hat_zoom} display the same for the shape in Figure \ref{fig:Lady}.}
\label{fig:algorithm_performance}
\end{figure*} 
\begin{figure}[]
\centering
\hfill
\begin{minipage}[t]{\linewidth}
\centering
\subfloat[]{\includegraphics[height=0.45\linewidth,valign=b]{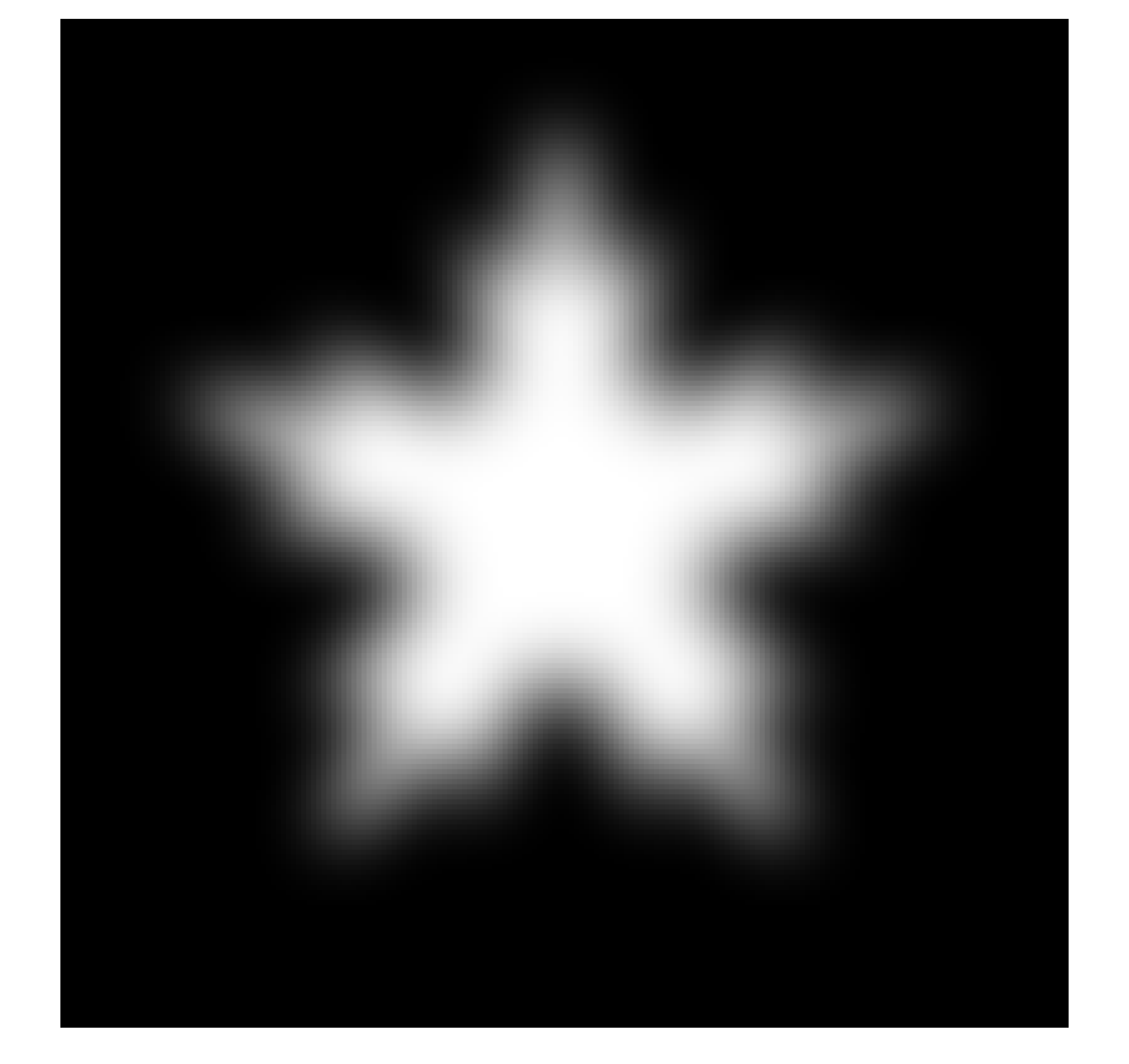}\label{fig:device_image_blurred}}
\end{minipage}
\hfill
\\
\hfill
\begin{minipage}[t]{\linewidth}
\centering
\subfloat[]{\includegraphics[height=0.39\linewidth]{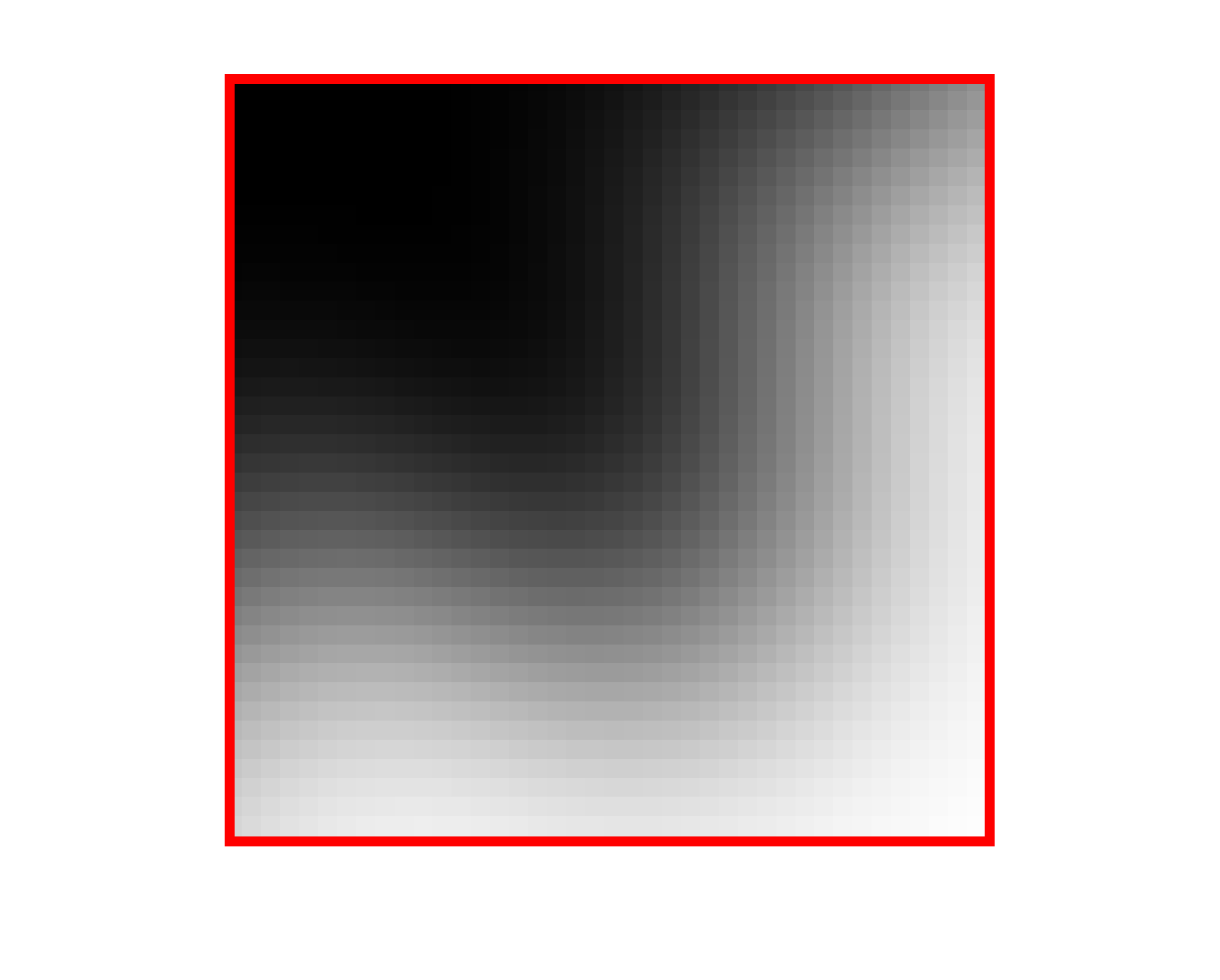}\label{fig:device_image_blurred_zoom}}
\subfloat[]{\includegraphics[height=0.39\linewidth]{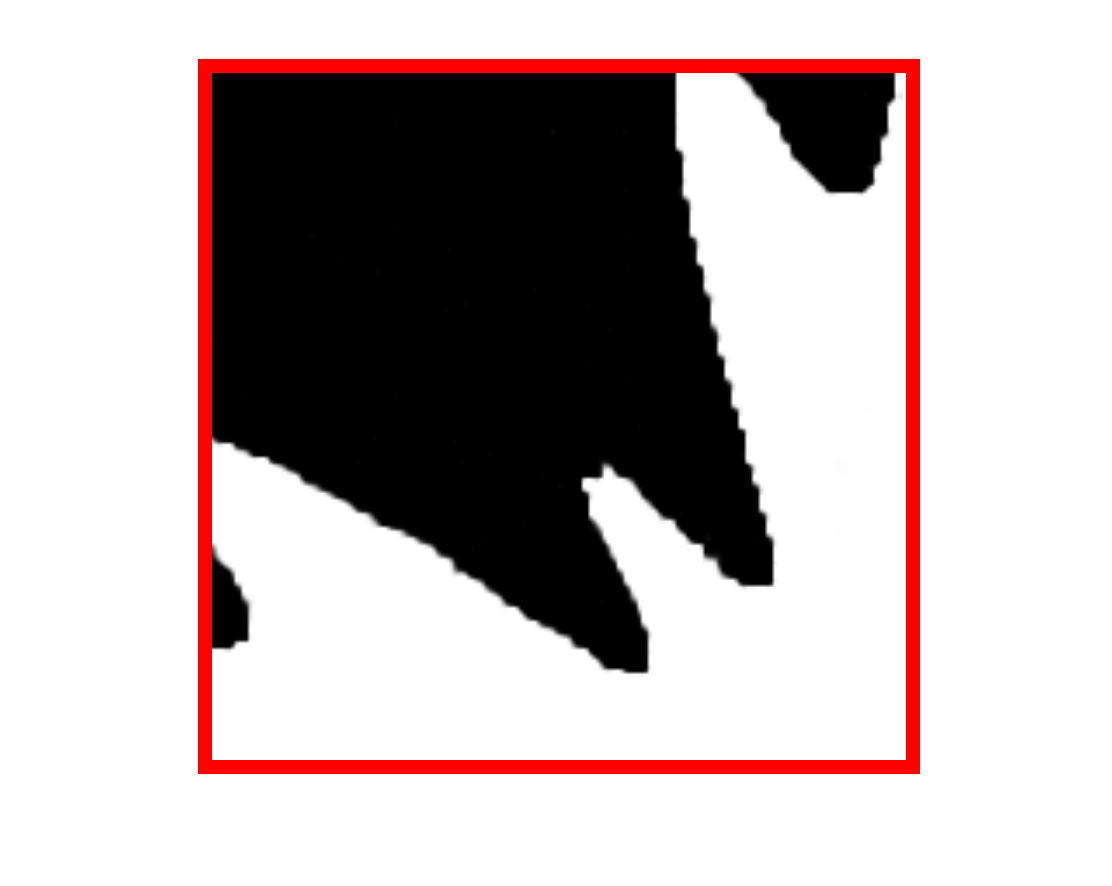}\label{fig:device_hat_blurred_zoom}}
\end{minipage}
\hfill
\caption{Consistent shape reconstruction with the proposed algorithm from highly blurred discrete images: Figure \ref{fig:device_image_blurred} shows a $200\times 200$ discrete image corresponding to the shape in Figure \ref{fig:device}, when the sampling kernels are shifts of an stretched biquadratic B-spline with an effective support of $40\times 40$ pixels. Figs. \ref{fig:device_image_blurred_zoom} and \ref{fig:device_hat_blurred_zoom} show the same enlarged sections (as in Figure \ref{fig:device_zoom}) of the measurement image and reconstructed shape, respectively. The recovered image (without any thresholding) has a PSNR of 33.8096 dB with respect to the original shape and a measurement PSNR of 75.0489 dB.}
\label{fig:algorithm_performance_blur}
\end{figure}
Next, we examine the solutions of \eqref{eq:P_1} to $200\times 200$-pixel discrete images of the shapes depicted in Figures \ref{fig:device} and \ref{fig:Lady} at resolution $1000\times 1000$ (Figure \ref{fig:Lady} is taken from the middle part of Figure \ref{fig:Matisse}). The sampling kernel for this experiment is the biquadratic B-spline. Figure \ref{fig:algorithm_performance} presents the same enlarged sections of the original shapes, the discrete images (just for a visual comparison) and their reconstructions with the proposed algorithm. The figures demonstrate that both reconstructed images are almost binary. Also, Table \ref{tab:table1} shows the quantitative evaluation of the reconstructed images. In this table, we also compare our results with the ones obtained by the interpolation of the measurement images with the bilinear B-spline kernel, followed by a thresholding at level 0.5. For a fair comparison, we also threshold our results to calculate the PSNRs. The numbers in this table clearly indicate the success of our proposed algorithm for consistent shape reconstruction.

\begin{table}
\begin{center}
\caption{Quantitative evaluation of the proposed algorithm (numbers in dB)}
\label{tab:table1}
\begin{tabular}{lc|c||c|c}
&\multicolumn{2}{c||}{shape in Figure \ref{fig:device}}&\multicolumn{2}{c}{shape in Figure \ref{fig:Lady}}\\
\cline{2-5}
&image&measurement&image&measurement\\
&PSNR&PSNR&PSNR&PSNR\\
\cline{1-5}
proposed &\multirow{2}{*}{29.1507}&\multirow{2}{*}{58.3316}&\multirow{2}{*}{43.8839}&\multirow{2}{*}{63.3429}\\
solution&&&&\\
\hline
linear &\multirow{2}{*}{26.4397}&\multirow{2}{*}{41.4224}&\multirow{2}{*}{33.7172}&\multirow{2}{*}{49.1746}\\
interpolation&&&&
\end{tabular}
\end{center}
\end{table}

For a given shape image, the resolution requirement in Definition \ref{defi:unifiable} mainly depends on the sampling grid, rather than the PSF. To examine this fact, we repeat the experiment in Figure \ref{fig:algorithm_performance} by regenerating a $200\times 200$ discrete image from Figure \ref{fig:device} using a stretched biquadratic B-spline sampling kernel with an effective support of $40\times 40$ pixels. The result is the highly blurred image in Figure \ref{fig:device_image_blurred}. Also, Figure \ref{fig:device_image_blurred_zoom} shows the enlarged section equivalent to Figure \ref{fig:device_zoom}. The quality of the reconstructed image in Figure \ref{fig:device_hat_blurred_zoom} ($\textrm{PSNR} = 33.8 \text{dB}$) confirms that the sampling grid outweighs the choice of the PSF in determining the performance.


\begin{figure*}[]
\centering
\subfloat[]{\includegraphics[width=0.30\linewidth]{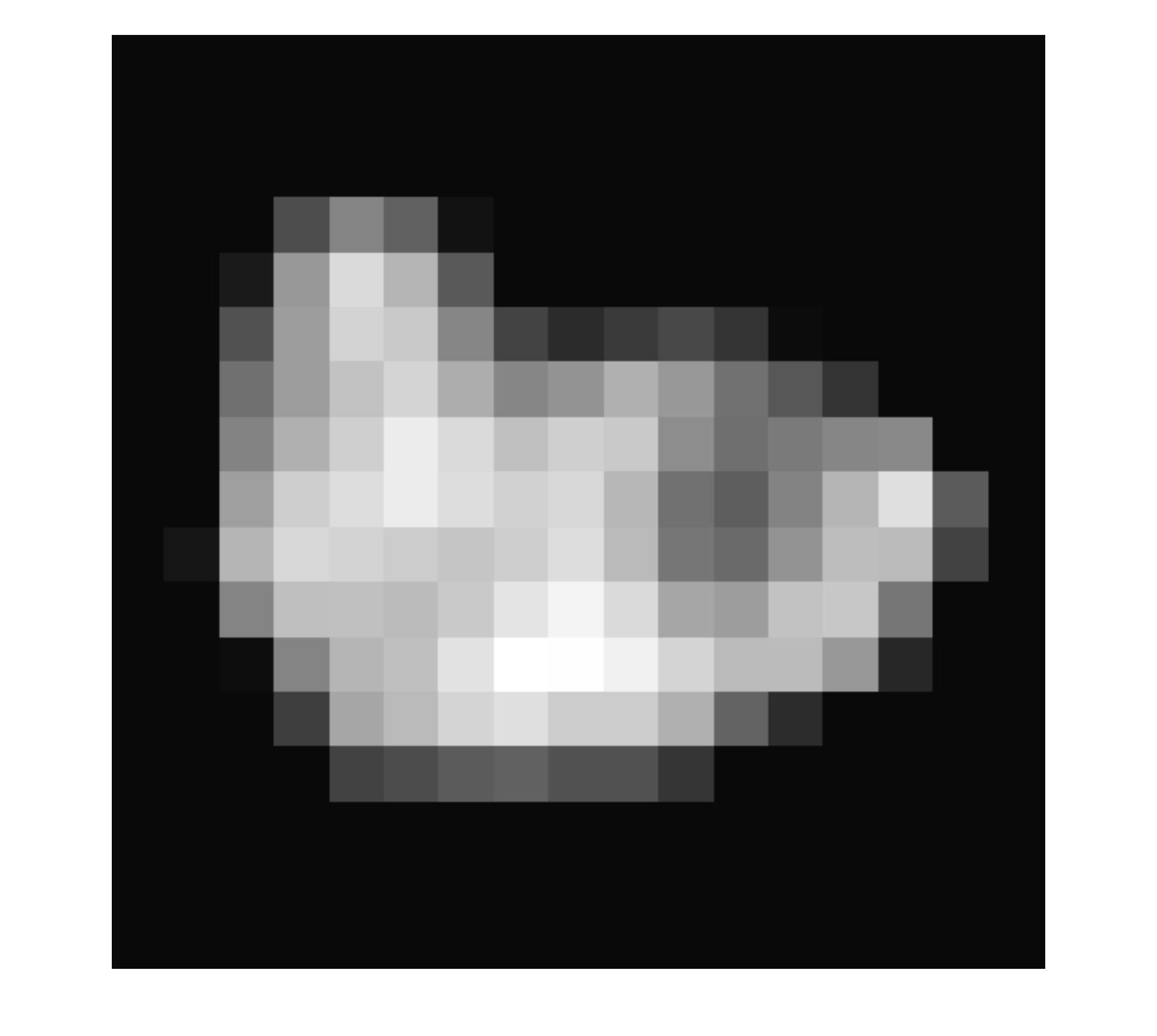}\label{fig:Hydra_samples}}
\hfill
\subfloat[]{\includegraphics[width=0.30\linewidth]{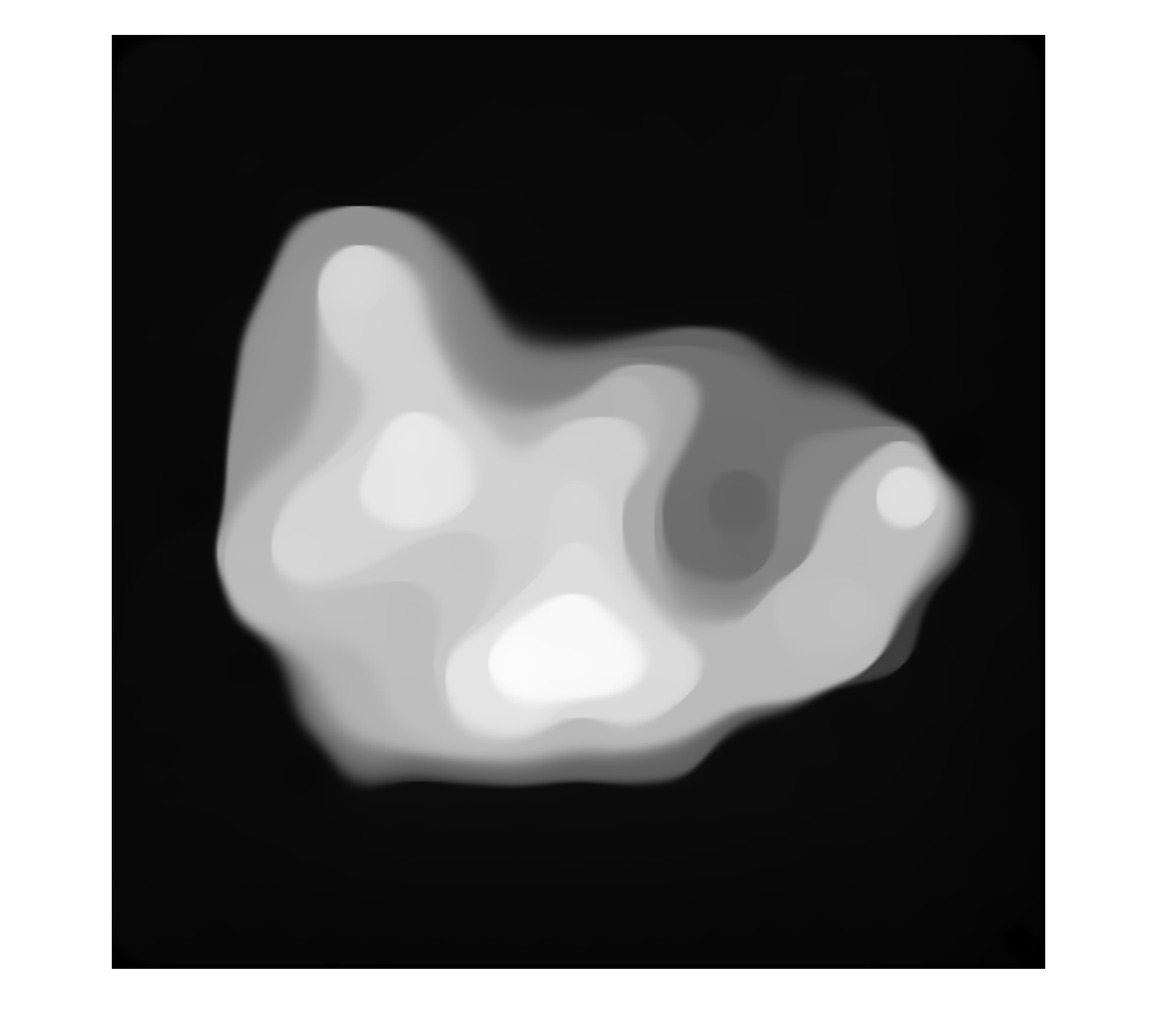}\label{fig:Hydra_processed}}
\hfill
\subfloat[]{\includegraphics[width=0.30\linewidth]{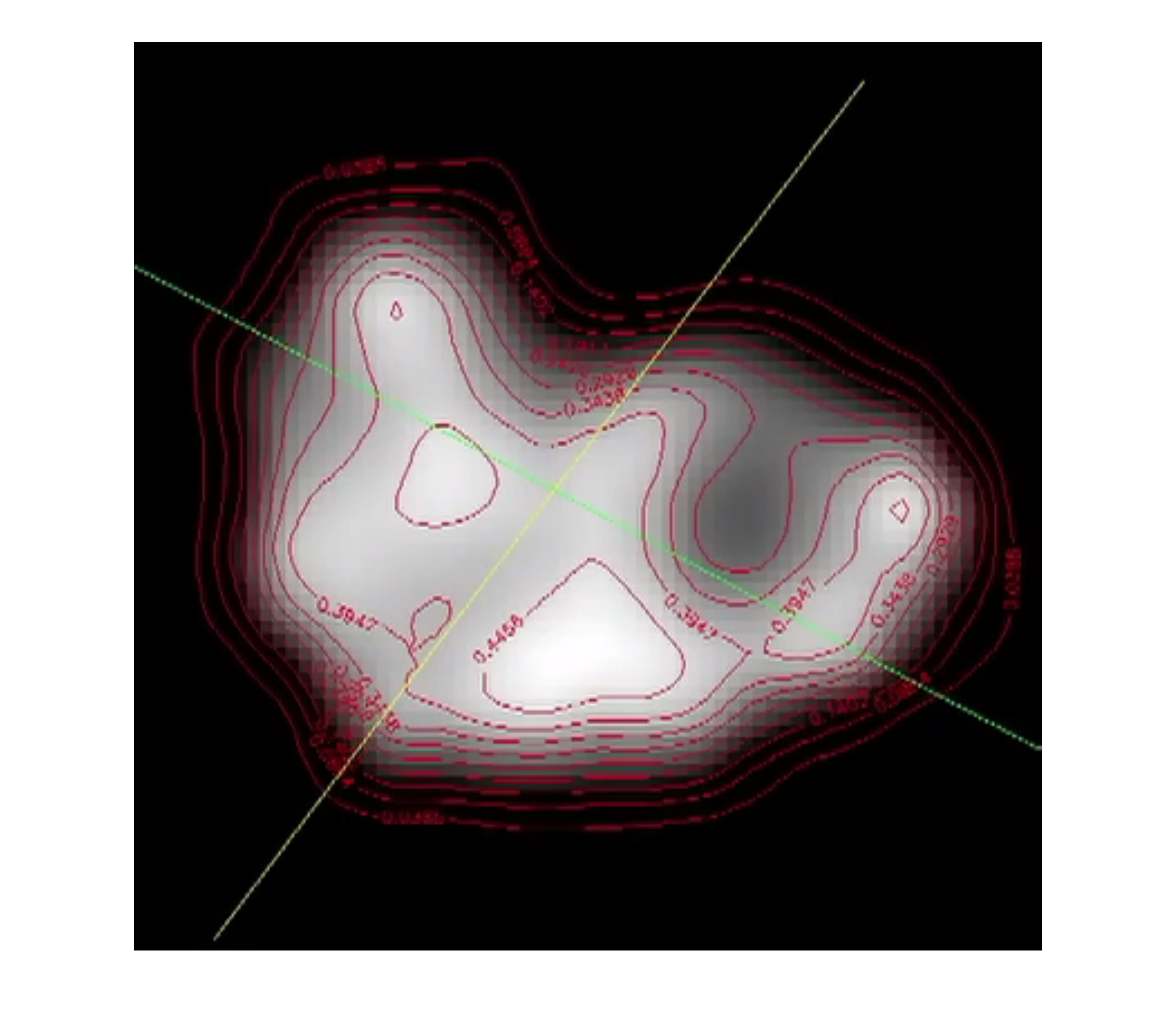}\label{fig:Hydra_processed_NASA}}

\caption{Performance of the proposed method in a setting with limited measurements: the measurement image in (a) is taken by the imager of the New Horizons space probe from Hydra. Due to the deficiency of measurements, our reconstruction  in (b) is not bilevel; yet it is a good match to the processed image (c) released by NASA.}
\label{fig:object}
\end{figure*}
Finally in the last experiment, we study the performance of the proposed method in a setting severely deficient in the number of measurements. For this purpose, we consider a recent image by the \emph{New Horizons} spacecraft in July 2015 from a moon of Pluto named \emph{Hydra}. Figure \ref{fig:Hydra_samples} depicts the received measurements. Although a high resolution imager is used, due to the long distance of the spacecraft to Hydra compared to the size of Hydra, we observe a highly pixelated image. According to the available data, the effective PSF width of the imager is around 1.5 pixels, which we model by a dilated biquadratic B-spline. 
Figure \ref{fig:Hydra_processed} shows the output of the convex program to the measurements by applying the approximate PSF. As the measurements are too few, the reconstructed image is not bilevel (indeed, it is not unlikely to assume the image of Hydra being binary from this distance). Nevertheless, it is interesting to note that the obtained multi-level image is not far from the processed image released by NASA in Figure \ref{fig:Hydra_processed_NASA}.

\section{Conclusion}\label{sec:conclusion}
In this paper, we studied the problem of reconstructing a continuous-domain shape image from the samples in a gray-scale discrete image. This is essentially equivalent to the interpolation of pixels in a way that generates a binary image. We formulated this problem as a minimization problem where the functional is the continuous-domain total variation and the constraints encode the sampling relation between the continuous-domain image and the pixels of the discrete image. When the search is over binary images, the minimizers will be shapes with minimum perimeter and smooth boundaries that satisfy the measurements. However, the search over shape images is computationally intractable. 
We introduced the reducibility condition on the samples of the discrete image and proved that when it is satisfied, extending the search domain to the non-negative-valued images would not omit any of the binary minimizers. The reducibility condition essentially calls for smooth changes in the values of the neighboring pixels. From this perspective, this is an intuitive requirement on the minimum sampling density that is needed for tracking local changes in the shape boundaries.

We conjecture that under the reducibility condition, the convex problem has a unique binary solution. Nevertheless, we introduced a test to verify whether an obtained solution to the convex minimization problem is binary. This test is mainly useful in the numerical calculation of the minimizers where the recovered solutions might not be precisely bilevel due to the numerical precision.

Our approach in this paper was based on minimization of the total variation, but all the results remain valid if we use a weighted total variation. A carefully designed weighting kernel might locally adjust the recovered shapes and lead to shapes with higher mean curvature.

\appendix[Proof of Lemma \ref{lem:mainLemma}]
\noindent In the following, we reserve the notation $\mathbf{e}^{n}_i$ for the canonical basis of $\R^n$:
\begin{align*}
\mathbf{e}^{n}_i \triangleq \big[0,\dots,0,\stackbin[i\rm{th}]{}{1},0,\dots,0\big]^T\in\R^n.
\end{align*}
We prove the lemma by induction on $n$. We set the basis of the induction on $n=1$. It is trivial to check that Condition (\ref{cod:linear_Comb}) for $n=1$ implies the claim in this case. Next, by assuming the validity of Lemma \ref{lem:mainLemma} for some $n\geq 1$, we demonstrate the validity for $n+1$. 

For the case $K_1^{(n+1)}=\emptyset$, it is not difficult to see that $\boldsymbol{\lambda}=[\tfrac{1}{n+1},\dots,\tfrac{1}{n+1}]^T$ satisfies the requirement. Here, Condition (\ref{cod:linear_Comb}) implies that all the inequalities of Condition (\ref{cond:equality}) are in fact equalities. Hence, we focus on $K_1^{(n+1)}\neq \emptyset$. Without loss of generality, we assume that $n+1\in K_1^{(n+1)}$. Next, we will try to reduce the $(n+1)$-dimensional problem into a similar $n$-dimensional one with $K_1^{(n)}=K_1^{(n+1)}\setminus \{n+1\}$ and $K_2^{(n)}=K_2^{(n+1)}$. 

According to Condition (\ref{cod:facets}), at $\boldsymbol{\lambda}=\mathbf{e}^{n+1}_{n+1}\in\Delta_{n+1}$ we have that
\begin{align*}
\forall\, i\in K_1^{(n+1)}\setminus\{n+1\}:~~v_i(\mathbf{e}^{n+1}_{n+1}) < d_i.
\end{align*}
If $K_1^{(n+1)}\setminus\{n+1\} = \emptyset$, set $\epsilon = \tfrac{1}{2}$. Otherwise, set $0<\epsilon \leq \tfrac{1}{2}$ such that for all $i\in K_1^{(n+1)}\setminus\{n+1\}$ and all $\Lam\in\Delta_{n+1}$ with  $\|\Lam-\mathbf{e}^{n+1}_{n+1}\|<\epsilon$ (\emph{i.e.}, $\epsilon$-neighborhood of $\mathbf{e}^{n+1}_{n+1}$ inside $\Delta_{n+1}$),  we have that $v_i(\Lam)<d_i$. The existence of such $\epsilon$ follows from the continuity of $v$ (and consequently $v_i$s). Furthermore, Condition (\ref{cond:equality}) implies $v_i(\Lam)\leq d_i$ for all $i\in K_2^{(n+1)}$ and the same set of $\boldsymbol{\lambda}$ vectors. In summary, we conclude the existence of $0<\epsilon \leq \tfrac{1}{2}$ such that
\begin{align*}
\forall\, 1\leq i\leq n,\, \Lam\in\Delta_{n+1}, \, \|\Lam-\mathbf{e}^{n+1}_{n+1}\|<\epsilon:~ v_i(\Lam)\leq d_i.
\end{align*}
By taking Condition (\ref{cod:linear_Comb}) into account, we observe that
\begin{align}\label{eq:neighbourhood}
\forall\, \Lam\in\Delta_{n+1},~ \|\Lam-\mathbf{e}^{n+1}_{n+1}\|<\epsilon:~~ v_{n+1}(\Lam) \geq  d_{n+1}.
\end{align}
In words, the value of $v_{n+1}$ in a neighborhood of $\mathbf{e}^{n+1}_{n+1}$ never drops below the desired value $d_{n+1}$. In contrast, the values of $v_{n+1}$ on the facet of the simplex $\Delta_{n+1}$ opposite to $\mathbf{e}^{n+1}_{n+1}$ ($\Lam\in\Delta_{n+1},~ \lambda_{n+1}=0$) are strictly below $d_{n+1}$ according to Condition (\ref{cod:facets}). Since $v_{n+1}$ is continuous, by starting from any point on this facet and gradually moving towards $\mathbf{e}^{n+1}_{n+1}$ on the line connecting the two points, $v_{n+1}$ will eventually attain the value $d_{n+1}$. By considering the points on all such lines that $v_{n+1}$ attains the value $d_{n+1}$ for the first time (when moving away from the facet towards the vertex $\mathbf{e}^{n+1}_{n+1}$), we shall have a manifold intersecting with all the facets except possibly the studied one. To mathematically represent this manifold we employ the following definition:
\begin{align}
\forall\,\mathbf{t}\in\Delta_n:~~ \beta(\mathbf{t})\triangleq \inf \Big\{\beta\in[0,1]~\Big|~ \forall\, \gamma, \,\beta\leq \gamma \leq 1:\nonumber\\~ v_{n+1}\big(\gamma\,t_1,\,\dots, \,\gamma\, t_{n},\, 1-\gamma\big)<d_{n+1} \Big\}.
\end{align}
It is not difficult to apply the continuity of $v_{n+1}$ to conclude the continuity of $\beta(\mathbf{t})$ and the fact that
\begin{align}\label{eq:fn_1}
\forall\,\mathbf{t}\in\Delta_n: ~~v_{n+1}\Big(\beta(\mathbf{t})t_1,\,\dots,\,\beta(\mathbf{t})t_n,\, 1-\beta(\mathbf{t})  \Big)= d_{n+1}.
\end{align}
Moreover, we invoke \eqref{eq:neighbourhood} to demonstrate that $\beta(\mathbf{t})\geq \frac{\epsilon}{\sqrt{2}}$; \emph{i.e.}, $\beta(\mathbf{t})$ is strictly positive for all $\mathbf{t}\in\Delta_n$. 

Now we are ready to reduce the dimension to $n$. For this purpose, we define the function $u:\Delta_n\mapsto \R^n$ as
\begin{align}\label{eq:hDef}
&\forall\,\mathbf{t}=[t_1,\dots,t_n]^T\in\Delta_n:\\
&u(\mathbf{t})\triangleq \left[\begin{array}{c}
v_1\Big(\beta(\mathbf{t})t_1,\,\dots,\,\beta(\mathbf{t})t_n,\, 1-\beta(\mathbf{t})  \Big)\\
\vdots \\
v_n\Big(\beta(\mathbf{t})t_1,\,\dots,\,\beta(\mathbf{t})t_n,\, 1-\beta(\mathbf{t})  \Big)
\end{array}\right] 
= \left[\begin{array}{c}
u_1(\mathbf{t}) \\
\vdots \\
u_n(\mathbf{t})
\end{array}\right].\nonumber
\end{align}
The continuity of $u(\mathbf{t})$ directly follows from the continuity of $v$ and $\beta$. To verify Condition (\ref{cod:linear_Comb}) for $u$ note that
\begin{align*}
&\sum_{i=1}^{n} \beta(\mathbf{t})t_i\, v_i\Big(\beta(\mathbf{t})t_1,\,\dots,\,\beta(\mathbf{t})t_n,\, 1-\beta(\mathbf{t})  \Big) \nonumber\\
&+ \big(1-\beta(\mathbf{t}) \big) \underbrace{v_{n+1}\Big(\beta(\mathbf{t})t_1,\,\dots,\,\beta(\mathbf{t})t_n,\, 1-\beta(\mathbf{t})  \Big)}_{d_{n+1}} \\
&= \sum_{i=1}^{n} \beta(\mathbf{t})t_i\, d_i + \big(1-\beta(\mathbf{t}) \big)d_{n+1}\\
& \stackrel{\beta(\mathbf{t})\neq 0}{\Longrightarrow}   \sum_{i=1}^{n} t_i\, \underbrace{v_i\Big(\beta(\mathbf{t})t_1,\,\dots,\,\beta(\mathbf{t})t_n,\, 1-\beta(\mathbf{t})  \Big)}_{u_i(\mathbf{t})} = \sum_{i=1}^{n} t_i\, d_i.
\end{align*}

Also, let $\mathbf{t}\in\Delta_n$ be such that $t_i=0$ for some $1\leq i\leq n$. Recalling the definition of $u$, we have that
\begin{align*}
u_i(\mathbf{t}) = v_i(\widetilde{\Lam}),
\end{align*}
where
\begin{align*}
\widetilde{\Lam} &= \Big[\beta(\mathbf{t})t_1,\,\dots,\,\beta(\mathbf{t})t_n,\, 1-\beta(\mathbf{t})  \Big]^T,\\
\sum_{i=1}^{n+1}\widetilde{\lambda}_i=&\beta(\mathbf{t})\underbrace{\sum_{i=1}^{n}t_i}_{=1}+1-\beta(\mathbf{t})=1~~ \Rightarrow ~~ \widetilde{\Lam}  \in \Delta_{n+1}.
\end{align*}
As $t_i=0$ results in $\widetilde{\lambda}_i=0$, the Conditions (\ref{cod:linear_Comb}) and (\ref{cond:equality}) directly carry over to the functions $u_i$ with $K_1^{(n)} = K_1^{(n+1)}\setminus\{n+1\}$ and $K_2^{(n)}=K_2^{(n+1)}$.

To sum up, $u$ is a continuous function that satisfies Conditions (\ref{cod:linear_Comb})-(\ref{cond:equality}). Therefore, we conclude by the assumption of the induction that there exists $\mathbf{t}^{*}\in\Delta_n$ such that
\begin{align*}
u(\mathbf{t}^{*}) = [d_1,\dots,d_n]^T. 
\end{align*}
Finally, by plugging this result into \eqref{eq:hDef} and using \eqref{eq:fn_1}, we obtain that
\begin{align*}
v\Big(\beta(\mathbf{t}^{*})t^{*}_1,\,\dots,\,\beta(\mathbf{t}^{*})t^{*}_n,\, 1-\beta(\mathbf{t}^{*})  \Big) = [d_1,\dots,d_{n+1}]^T.
\end{align*}
 $\hspace{\stretch{1}}\blacksquare$

\bibliographystyle{IEEEtran}
\bibliography{refs}

\ifCLASSOPTIONcaptionsoff
  \newpage
\fi

\end{document}